\RequirePackage{fix-cm}
\documentclass[twocolumn]{svjour3}          
\smartqed  

\usepackage{graphicx}
\usepackage{amsfonts}
\usepackage{amsmath}
\usepackage{amssymb}
\usepackage{graphics}
\usepackage{epsfig}
\usepackage{float} 

\newcommand{\say}[1]{``#1''}
\newcommand{\f}[1]{\mathbf{#1}}

\newcommand{\m}[1]{\mathcal{#1}}

\newcommand{\dv}{\operatorname{div}}
\newcommand{\tr}{\operatorname{tr}}

\newcommand{\sg}{\hspace{0.25cm}}
\newcommand{\bg}{\hspace{0.5cm}}

\journalname{Computational Mechanics}

\begin{document}

\title{Mesh deformation techniques in fluid-structure interaction: robustness, accumulated distortion and computational efficiency
}

\author{Alexander Shamanskiy         \and
        Bernd Simeon 
}

\institute{A. Shamanskiy \at
              TU Kaiserslautern, Department of Mathematics, Felix-Klein-Zentrum, Paul-Ehrlich-Str. 31, 67663, Kaiserslautern, Germany \\
              Tel.: +49 631 205 5318\\
              \email{shamansk@mathematik.uni-kl.de}\\
              ORCID: 0000-0003-3580-3332        
           \and
           B. Simeon \at
              TU Kaiserslautern, Department of Mathematics, Felix-Klein-Zentrum, Paul-Ehrlich-Str. 31, 67663, Kaiserslautern, Germany
}

\date{Received: date / Accepted: date}

\maketitle

\begin{abstract}
 An important ingredient of any moving-mesh method for fluid-structure interaction (FSI) problems is the mesh deformation technique (MDT) used to adapt the computational mesh in the moving fluid domain. An ideal technique is computationally inexpensive, can handle large mesh deformations without inverting mesh elements and can sustain an FSI simulation for extensive periods ot time without irreversibly distorting the mesh. Here we compare several commonly used techniques based on the solution of elliptic partial differential equations, including harmonic extension, bi-harmonic extension and techniques based on the equations of linear elasticity. Moreover, we propose a novel technique which utilizes ideas from continuation methods to efficiently solve the equations of nonlinear elasticity and proves to be robust even when the mesh is subject to extreme deformations. In addition to that, we study how each technique performs when combined with the Jacobian-based local stiffening. We evaluate each technique on a popular two-dimensional FSI benchmark reproduced by using an isogeometric partitioned solver with strong coupling.
\keywords{ALE methods \and isogeometric analysis \and nonlinear elasticity \and continuation methods \and local stiffening}
\end{abstract}

\section{Introduction}
Fluid-structure interaction (FSI) constitutes a class of problems involving two-way dependence between structural objects and a fluid. As such, FSI is a vast topic with applications spanning a  spectrum from aerospace and civil engineering \cite{bazilevs2013} to biomechanical and cardiovascular simulations \cite{formaggia2010cardiovascular}. In FSI problems, the fluid exerts a force on the structure which deforms in response. As the structure moves, it changes the shape of the domain occupied by the fluid together with the fluid motion and, as a result, the force that the fluid exerts on the structure. This two-way coupling between the fluid and structure behavior as well as the necessity to accommodate the fluid domain motion in both the continuous and discrete formulations of the problem is what makes FSI so notoriously complex. 

Since FSI problems rarely admit analytical solutions, computational methods are widely adopted in FSI research. Here, one can distinguish between the static-mesh and moving-mesh methods. While the former attempt to resolve the motion of the fluid domain implicitly, for example by means of a stationary background Cartesian mesh \cite{schillinger2012isogeometric}, the latter deal with the motion of the fluid domain by tracking its boundary and adapting the computational mesh correspondingly. In this work, we study and compare various mesh deformation techniques (MDTs) which can be used to adapt the fluid mesh if the moving-mesh methods are used. The main focus here lies on the robustness of a given technique, that is, how much mesh distortion it introduces and how much mesh deformation it can handle without entangling or inverting mesh elements. Additionally, we pay attention to computational complexity of MDTs, which can significantly increase the overall FSI simulation time if left unchecked.

To study performance of MDTs in their natural habitat, we employ a popular two-dimensional FSI benchmark from \cite{turek2006proposal}. In the benchmark, an unstable flow of a viscous fluid leads to development of the vortex shedding phenomenon which results in oscillations of a flexible beam structure. The oscillations grow in magnitude until they reach a stable periodic regime which lends itself well to studying possible long-term effects of MDTs on the fluid mesh. One of such long-term effects is the accumulated mesh distortion  where mesh elements increasingly become permanently distorted,  deteriorating the simulation accuracy. In addition to the original benchmark, we employ its simplified version with no fluid mechanics involved to perform a large number of computationally inexpensive tests. In this way, we can concentrate on mesh deformation and conduct a detailed analysis of MDT behavior. 

To reproduce the FSI benchmark, we use a partitioned solver with strong coupling and Aitken relaxation \cite{kuttler2008fixed}. Although modern space-time (ST) methods are becoming increasingly common in FSI \cite{bazilevs2013,tezduyar2007modelling,takizawa2012space}, we resort to classical arbitrary Lagrangian-Eulerian (ALE) methods \cite{bazilevs2013,richter2010finite,hughes1981lagrangian} which are more straightforward in implementation. Our choice in favor of a basic partitioned ALE solver is justified since both ST and ALE methods make use of the same MDTs, so we can focus on mesh deformation in this paper.

All MDTs we consider here are based on solution of elliptic partial differential equations. These include existing techniques such as harmonic extension  \cite{dorfel2011fluid,wu2014fully}, bi-harmonic extension \cite{richter2010numerical} and a widely adopted technique based on linear elasticity theory \cite{tezduyar1992computation,johnson1994mesh}. Moreover, we propose an efficient MDT based on the equations of nonlinear elasticity and a logarithmic neo-Hookean material law which we refer to as tangential incremental nonlinear elasticity (TINE). Although techniques based on the equations of nonlinear elasticity have been proposed before \cite{takizawa2013fluid,suito2014fsi}, TINE is novel in that it uses the idea of a tangential continuation method \cite{deuflhard2011newton,shamanskiy2019mesh} to efficiently solve the corresponding nonlinear equations. As a result, TINE is only slightly more computationally expensive than the linear-elasticity-based techniques which are linear in nature. On the other hand, it can handle as much mesh deformation but does not suffer from the accumulated distortion effect.

Robustness of any MDT can be increased by additional augmentations. Probably the most popular one is the Jacobian-based local stiffening \cite{tezduyar1993parallel,stein2003mesh} which turns individual mesh elements stiffer or softer depending on their size and shape. In this work, we study how different MDTs react to the Jacobian-based local stiffening. Although not considered here, further possible MDT augmentations include solid layer extension \cite{tezduyar2001finite} and element relaxation \cite{takizawa2013fluid,takizawa2020low}.

The research we present in this work has been conducted in the framework of isogeometric analysis \cite{Hughes2005,Cottrell.2009}. Despite that, the results are applicable to classical finite elements methods or any other mesh-based method for solving partial differential equations.

The rest of this paper is structured as follows. Section \ref{chap:benchmark} outlines geometry and settings of the FSI benchmark and fixes the necessary notation. In Section \ref{chap:methods}, we describe various MDTs considered in this work as well as the Jacobian-based local stiffening. In Section \ref{chap:testALE}, we consider the simplified benchmark and conduct a detailed analysis of the short-term and long-term behavior of different MDTs in artificial FSI-like conditions. After that, we proceed to performing a full FSI simulation of the benchmark in Section \ref{chap:testFSI}. We study the performance of the MDTs and check if the choice of a particular MDT affects the simulation results. Finally, we discuss the results of the MDT analysis and FSI simulations, draw conclusion and outline further research directions in Section \ref{chap:concl}.   

\section{Benchmark description}\label{chap:benchmark}
\begin{figure*}
	\centering
	\includegraphics[height=3cm]{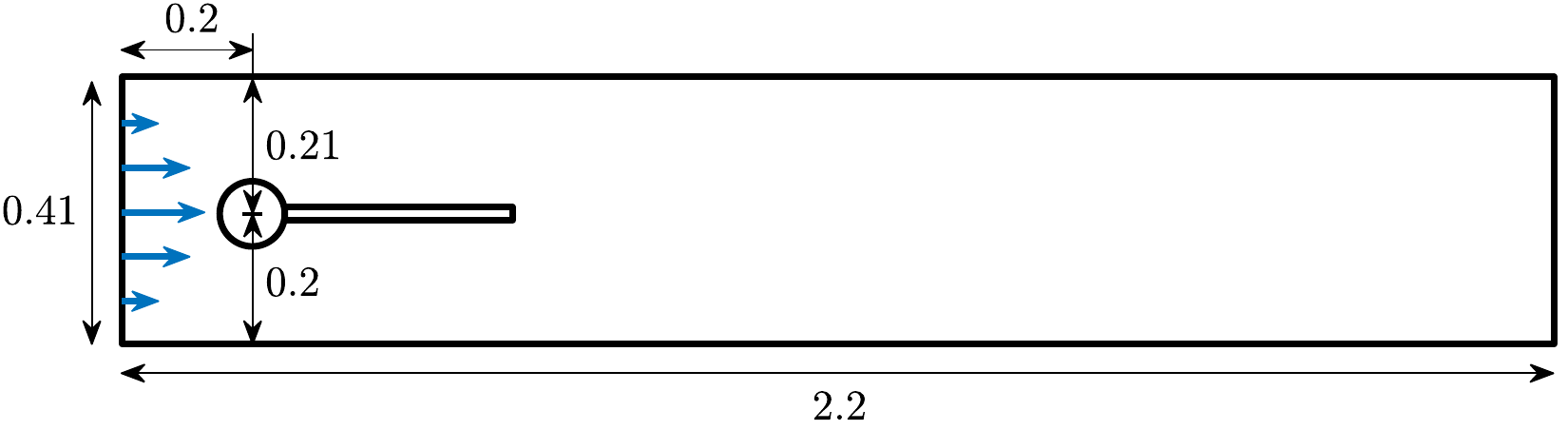}\\
	\includegraphics[height=2cm]{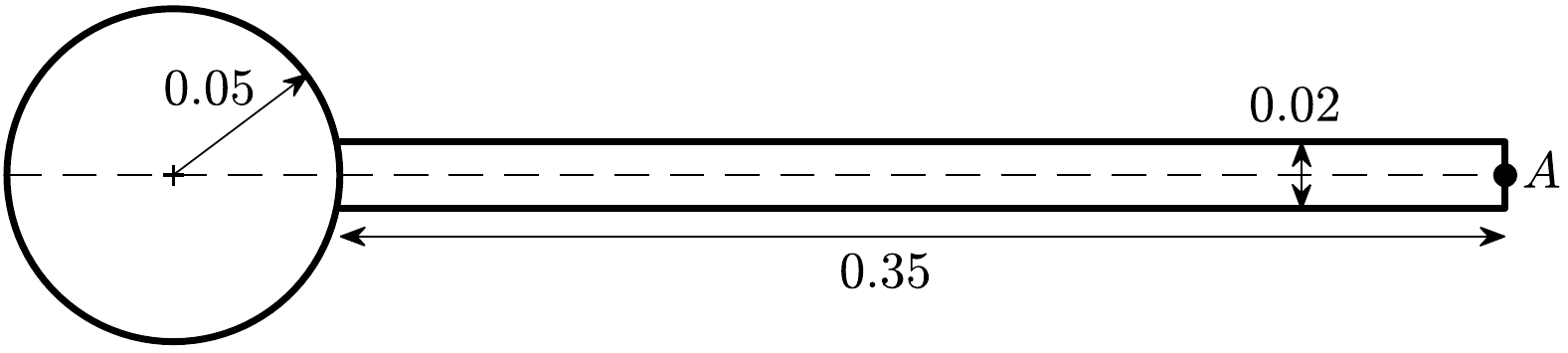}
	\caption{Top: the flow channel and the submerged structure in the initial configuration. Bottom: close-up on the structure.}
	\label{fig:geometry}
\end{figure*}

In this section, we describe the geometry and mathematical model of the FSI benchmark from \cite{turek2006proposal}.
It studies the flow of an incompressible Newtonian fluid through a 2D channel as the fluid interacts with a submerged structure. The channel is a rectangle $[0,2.2]\times[0,0.41]$. The structure consists of a rigid disk $B_{0.05}(0.2,0.2)$ and a flexible beam $[0.2,0.6]\times[0.19,0.21]\backslash B_{0.05}(0.2,0.2)$ which is attached at its left end to the boundary of the disk. Figure \ref{fig:geometry} illustrates the setting; note that the geometry is intentionally non-symmetric. 

The top and bottom walls of the channel are impermeable to the fluid. The fluid enters the channel through the left wall with a prescribed velocity and exits through the right wall freely. The presence of the submerged structure changes the flow of the fluid and, in response, the fluid exerts a certain force on the structure. Thus, this is an FSI problem. Depending on the prescribed material parameters of the fluid and beam as well as on the inflow fluid velocity, this fluid force can result in a noticeable deformation of the beam. The beam deformation alters the shape of the flow channel, the flow itself and, as a consequence, the force exerted on the structure by the fluid. Such systems with a two-way interaction between components are called coupled systems. 

In what follows, when we use the word \say{structure}, we refer to the flexible beam only. If the rigid disk is also considered, we explicitly mention it. Throughout this work, we use subscripts $s$ and $f$ to distinguish between  objects related to the structure and fluid respectively. Moreover, we use subscript $a$ when dealing with ALE mappings and related objects. Let $\Omega_f(t) \subset \mathbb{R}^2$ and $\Omega_s(t) \subset \mathbb{R}^2$ denote domains occupied by the fluid and structure respectively at time $t\in[0,T]$. Additionally, we use $\Omega^0_f=\Omega_f(0)$ and $\Omega^0_s=\Omega_s(0)$ to denote the fluid and structure domains at time $t=0$. Furthermore, we use $\Gamma(t)$ to denote the FSI interface $\partial\Omega_f(t)\cap\partial\Omega_s(t)$ where the interaction between the fluid and structure takes place. Correspondingly, $\Gamma^0=\Gamma(0)$ denotes the FSI interface at time $t=0$. In the rest of this section, we briefly formulate the equations that we use to describe the motion of the beam and the fluid as well as their interaction. The main goal here is to fix the notation necessary for the ensuing description of mesh deformation techniques (MDTs). We encourage the reader to consult the following references for more information on: finite element discretization for the Navier-Stokes equations for incompressible flows \cite{john2016finite}; finite element discretization for nonlinear elasticity problems \cite{wriggers2008nonlinear,bernal2013isogeometric}; ALE formulation of FSI problems \cite{richter2010numerical,bazilevs2013}; partitioned solution approach to FSI problems \cite{kuttler2008fixed}.

\subsection{Structure motion}
We assume that the structure behavior can be characterized as elastic and compressible. Let the structure in its initial (undeformed) configuration occupy the domain $\Omega_s^0$. We can describe the structure motion in terms of a displacement field $\f{u}_s: \Omega^0_s\times[0,T] \to \mathbb{R}^2$. Note that displacement $\f{u}_s$ can describe a rigid body motion with no deformation involved. Information on whether actual deformation of $\Omega_s^0$ takes place is contained in the deformation gradient $\f{F}_s = \f{I} + \nabla\f{u}_s$. Two important objects derived from the deformation gradient are the Green-Lagrange strain tensor $\f{E}_s = (\f{F}_s^T\f{F}_s-\f{I})/2$ and the Jacobian determinant $J_s=\det\f{F}_s$.

For the material behavior, we use the St.~Venant-Kirchhoff constitutive law which links the second Piola-Kirchhoff stress tensor $\f{S}_s$ to the Green-Lagrange strain tensor $\f{E}_s$:
\begin{equation}\label{eq:StVK}
\f{S}_s = \lambda_s\tr(\f{E}_s)\f{I} + 2\mu_s\f{E}_s.
\end{equation}
The material law (\ref{eq:StVK}) includes the Lam\'e parameters $\lambda_s$ and $\mu_s$, which are constitutive parameters describing physical properties of the material. They can be computed from Young's modulus $E_s$ and Poisson's ration $\nu_s$ as
\begin{equation}
\lambda_s = \frac{\nu_s E_s}{(1+\nu_s)(1-2\nu_s)}\sg\text{and}\sg\mu_s = \frac{E_s}{2(1+\nu_s)}.
\end{equation}
Second Piola-Kirchhoff stress tensor $\f{S}_s$ measures forces appearing in the deformed structure with respect to its initial configuration $\Omega_s^0$. In FSI applications, it is important to have an ability to express these forces with respect to the deformed configuration $\Omega_s(t)$. This can be achieved by means of the Cauchy stress tensor $\pmb{\sigma}_s$, which is related to $\f{S}_s$ by
\begin{equation}
\pmb{\sigma}_s = \frac{1}{J_s}\f{F}_s\f{S}_s\f{F}_s^T.
\end{equation}

In the presence a given external acceleration $\f{g}:\Omega^0_s\times[0,T]\to\mathbb{R}^2$, the displacement $\f{u}_s$ should satisfy the local conservation equations of linear momentum
\begin{equation}\label{eq:beamBalance}
\rho_s\ddot{\f{u}}_s = \dv\f{P}_s + \rho_s\f{g} \text{ in }\Omega_s^0.
\end{equation}
Here, $\rho_s$ is the structure density which we assume to be constant in $\Omega_s^0$, and $\f{P}_s = \f{F}_s\f{S}_s$ is the first Piola-Kirchhoff stress tensor. Note that equations (\ref{eq:beamBalance}) are formulated in the stationary domain $\Omega_s^0$.

\subsection{Fluid motion}
To describe the fluid motion, we use the Navier-Stokes equations for incompressible flows. If a computational domain $\Omega_f$ does not change with time, the Navier-Stokes equations have the following form:
\begin{align}
\rho_f\dot{\f{v}}_f + \rho_f\nabla\f{v}_f\cdot\f{v}_f&=\dv\pmb{\sigma}_f + \rho_f\f{g},\\
\dv\f{v}_f &= \f{0} \text{ in }\Omega_f.
\end{align}
Here, $\f{v}_f:\Omega_f\times[0,T]\to\mathbb{R}^2$ is a vector field describing the fluid velocity at a given point in $\Omega$, $\rho_f$ is a constant fluid density, and $\pmb{\sigma}_f$ denotes a Cauchy stress tensor. Behavior of an incompressible Newtonian fluid is characterized by the following constitutive law: 
\begin{equation}
\pmb{\sigma}_f = -p_f\f{I} + \rho_f\nu_f(\nabla\f{v}_f + \nabla\f{v}_f^T),
\end{equation}
where $p_f:\Omega_f\times[0,T]\to\mathbb{R}$ is a pressure field, and $\nu_f$ denotes the kinematic viscosity of the fluid.

In FSI applications, one has to account for the motion of the fluid domain. In this work, we consider a common strategy based on the arbitrary Lagrangian-Eulerian (ALE) mappings. An ALE mapping describes the motion of the fluid domain in terms of an auxiliary displacement field $\f{u}_a:\Omega_f^0\times[0,T]\to\mathbb{R}^2$ such that $\Omega_f(t) = \Omega^0_f + \f{u}_a(\cdot,t)$. With an ALE mapping, the Navier-Stokes equations can be formulated in the moving domain $\Omega_f(t)$:
\begin{align}
\label{eq:NS_ALE1}\rho_f\dot{\f{v}}_f + \rho_f\nabla\f{v}_f\cdot(\f{v}_f-\dot{\f{u}}_a)&=\dv\pmb{\sigma}_f+ \rho_f\f{g},\\
\label{eq:NS_ALE2}\dv\f{v}_f &= \f{0} \text{ in }\Omega_f(t).
\end{align}

Since the fluid domain motion is driven by the structure deformation, the ALE displacement $\f{u}_a$ has to comply with the structure displacement $\f{u}_s$ on the FSI interface:
\begin{equation}
\f{u}_a = \f{u}_s\text{ on }\Gamma(t).
\end{equation}
Inside the fluid domain, $\f{u}_a$ can be chosen arbitrary, hence the term ALE. The only condition is that the ALE displacement should define an invertible deformation of the fluid domain. This means that the following condition has to hold:
\begin{equation}\label{eq:bijectivity}
J_a = \det\f{F}_a = \det(\f{I}+\nabla\f{u}_a) > 0.
\end{equation}
The scope of this work is to compare different options for defining the ALE displacement in the fluid domain provided the structure displacement on the FSI interface.

\subsection{Interaction conditions}
Physical interaction between the fluid and structure takes place on the FSI interface $\Gamma(t)$. It is characterized by the following two coupling conditions: the kinematic continuity
\begin{equation}
\f{v}_f = \dot{\f{u}}_s \text{ on }\Gamma(t),
\end{equation}
which assures that the fluid stays attached to the structure; and the dynamic continuity
\begin{align}
\nonumber&\pmb{\sigma}_s\cdot\f{n} = \pmb{\sigma}_f\cdot\f{n} \text { on }\Gamma(t) \sg\Leftrightarrow\\
&\f{P}_s\cdot\f{n}= J_a\pmb{\sigma}_f\f{F}_a^{-T}\cdot\f{n} \text{ on }\Gamma^0,
\end{align}
which maintains the balance of forces on the FSI interface. If a partitioned approach to FSI is used, one has to enforce the coupling conditions by exchanging information between the fluid and structure solvers.

\subsection{Initial and boundary conditions}
To complete the definition of the benchmark, we provide suitable initial and boundary conditions. The system is initialized with zero initial conditions for the fluid velocity, structure displacement and structure velocity:
\begin{align}
\nonumber&\f{v}_f(\cdot,0) = \f{0} \text{ in }\Omega^0_f,\\
&\f{u}_s(\cdot,0)=\f{0}\text{ in }\Omega_s^0,
\bg\dot{\f{u}}_s(\cdot,0)=\f{0}\text{ in }\Omega_s^0.
\end{align}
The main energy source of the system is an inflow boundary condition on the fluid velocity prescribed on the left end of the channel $\partial\Omega^\text{in}_f$. The condition prescribes a parabolic velocity profile
\begin{equation}
\f{v}_\text{par} = \left(\begin{array}{c}
v_{\max}\big(\frac{2}{0.41}\big)^2y(0.41-y)\\0\end{array}\right)
\end{equation}
with the maximum inflow velocity $v_\text{max}$ serving as an adjustable parameter. In order to comply with the initial conditions, $\f{v}_\text{par}$ is made time-dependent by scaling it with time:
\begin{equation}
\f{v}_\text{in}(t) = \left\{\begin{array}{c}
\f{v}_\text{par}\frac{1-\cos(\pi t/2)}{2}\;\sg\text{if }t<2\text{s},\\
\f{v}_\text{par}\sg\bg\bg\bg\sg\text{if }t\geqslant 2\text{s}.\end{array}\right.
\end{equation}
The resulting time-dependent inflow boundary condition $\f{v}_f = \f{v}_\text{in}$ provides a smooth warm-up phase for the simulation. On the right wall of the channel $\partial\Omega^\text{out}_f$, a do-nothing condition $\pmb{\sigma}_f\cdot\f{n} = \f{0}$ is prescribed. Additionally, a no-slip boundary condition $\f{v}_f = \f{0}$ is set on the rest of the channel wall $\partial\Omega^\text{ns}_f=\partial\Omega_f^0\setminus\partial\Omega^\text{in}_f\setminus\partial\Omega^\text{out}_f\setminus\Gamma^0$. Finally, the structure is fixed ($\f{u}_s=\f{0}$) on its left end $\partial\Omega_s^0\setminus\Gamma^0$.

\subsection{Geometry parametrization}
\begin{figure*}
	\centering
	\includegraphics[width=0.75\textwidth]{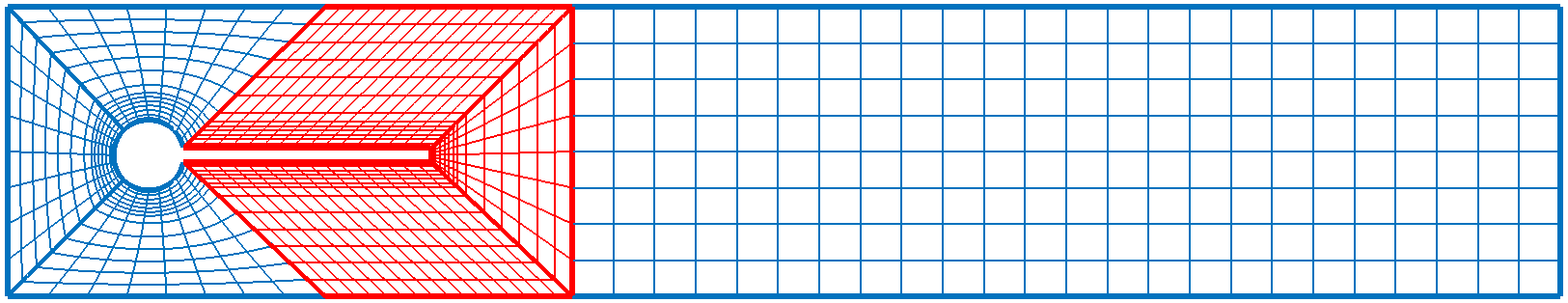}\\
	\caption{Computational mesh in the fluid domain. Three patches adjacent to the beam are chosen for ALE deformation.}
	\label{fig:flowMesh}
\end{figure*}

The research we present in this work has been conducted in the framework of isogeometric analysis \cite{Hughes2005,Cottrell.2009}.  In the spirit of IGA, we model computational domains as collections of tensor-product NURBS patches \cite{Piegl1997}. For the structure, we use a single quadratic NURBS patch, whereas the fluid domain is modeled with seven patches. Figure \ref{fig:flowMesh} illustrates the isoparametric mesh in the fluid domain after several applications of uniform $h$-refinement. For simplicity, we use matching parametrizations for neighboring fluid patches as well as for the structure patch and the surrounding fluid patches. This choice does not restrict the applicability of our results; however, it significantly simplifies the exchange of coupling information between the fluid and structure solvers.

Let us consider one of the NURBS patches forming the initial configuration of the fluid domain. Its parametrization $\f{G}_f^0:[0,1]^2\to\Omega_f^0$ can be written in terms of control points $\f{c}_k\in\mathbb{R}^2$ and tensor-product NURBS basis functions $N_k:[0,1]^2\to\mathbb{R}$ as
\begin{equation}
\f{G}_f^0(\xi,\eta) = \sum_{k=1}^n\f{c}_k N_k(\xi,\eta).
\end{equation}
In IGA, we the same NURBS basis functions to approximate solutions to PDEs. In our case, we use $N_k$ to discretize the ALE displacement $\f{u}_a$ in space. Assuming that we use time moments $t_0=0,\dots,t_i,\dots,t_N = T$ for the time discretization, we denote the ALE displacement at time $t_i$ by $\f{u}_a^i$. Its NURBS representation can be written as   
\begin{equation}
\f{u}_a^i(\xi,\eta) = \sum_{k=1}^n\f{d}_k^i N_k(\xi,\eta),
\end{equation}
where $\f{d}_k^i$ are the corresponding control points. Given $\f{u}_a^i$, we can easily construct a NURBS parametrization $\f{G}_f^i$ of the deformed fluid domain $\Omega_f^i=\Omega_f(t_i) = \Omega_f^0 + \f{u}_a^i$ as
\begin{equation}\label{eq:deformDomain}
\f{G}_f^i(\xi,\eta) =  \sum_{k=1}^n(\f{c}_k+\f{d}_k^i)N_k(\xi,\eta).
\end{equation}
Now $\f{G}_f^i$ can be used to discretize the Navier-Stokes equations (\ref{eq:NS_ALE1}--\ref{eq:NS_ALE2}) in the deformed configuration of the fluid domain. 

As a final comment, let us mention that it is possible to deform only a portion of the fluid domain while keeping the rest fixed. An obvious advantage of this approach is a considerable reduction in computational cost associated with construction of ALE mappings. In this work, we choose to deform only three patches of the fluid domain that are directly adjacent to the FSI interface. In Figure \ref{fig:flowMesh}, they are highlighted in red. In our experience, these three patches are more than capable of absorbing deformation of the structure that appears in this benchmark.

\section{Mesh deformation techniques}\label{chap:methods}
The scope of this paper is to compare various techniques to construct ALE mappings. Put simply, an ALE mapping is nothing else but a deformation of the computational mesh in the fluid domain. In this section, we introduce several commonly used mesh deformation techniques (MDTs) as well as propose certain variations to them. All considered techniques achieve the same goal: provided a displacement of the FSI interface at a given time, they extend it into the fluid domain. Note that although we present these techniques in a 2D setting, one can readily apply them in 3D. When comparing different techniques, we largely pay our attention to the maximum amount of deformation a particular technique can handle. That is, how much the mesh can be deformed before the bijectivity condition (\ref{eq:bijectivity}) is violated. A secondary measure is of course the overall computational cost associated with computing ALE mappings using a given MDT.

After MDTs, we describe the Jacobian-based local stiffening \cite{stein2003mesh} which we use to augment each of the considered techniques. Finally, we discuss practical ways to check whether the bijectivity condition (\ref{eq:bijectivity}) is satisfied.

\subsection{Harmonic extension (HE/IHE)}
Probably the simplest way to extend displacement of the FSI interface into the fluid domain is by means of harmonic extension (HE) \cite{richter2010numerical,dorfel2011fluid,wu2014fully}. Given the interface displacement $\f{u}_s^i$ at time $t_i$, the ALE displacement $\f{u}_a^i$ is computed by solving Laplace's equation in the initial configuration of the fluid domain $\Omega_f^0$ for every displacement component:
\begin{align}
\Delta\f{u}_a^i &= \f{0}\text{ in }\Omega_f^0,\\
\f{u}_a^i &= \f{u}_s^i\text{ on }\Gamma^0,\\
\f{u}_a^i &= \f{0}\text{ on }\partial\Omega_f^0\setminus\Gamma^0.
\end{align}
The interface displacement $\f{u}_s^i$ serves as a Dirichlet boundary condition on the FSI interface $\Gamma^0$. At the rest of the boundary, the prescribed displacement is zero. Note that the method does not take into account information about the interface or ALE displacement from the previous time step.

The HE technique is the least computationally expensive method to construct ALE mappings. Let $N$ be the number of inner control points in the fluid domain. Since all displacement components satisfy the same equation, they can be computed by solving a single linear system with an $N\times N$ matrix and an $N\times d$ right-hand side. Here, $d$ is the problem dimension; in our case, $d=2$. Moreover, once the matrix is assembled, it can be reused for all time steps, which drastically reduces the computational cost associated with matrix assembly in IGA. At each time step, one only has to update the right-hand side to take the current FSI interface displacement into account. This can be performed efficiently by storing a Dirichlet elimination matrix when assembling the main matrix.

Despite its computational efficiency, the HE technique has serious disadvantages. First of all, it treats displacement components as completely independent variables and does not promote bijectivity of the ALE mapping in any way. Second, solutions of Laplace's equation in the vicinity of corners behave like $r^{\pi/\omega}$, where $r$ is the distance to the corner and $\omega$ is the corner angle. For reentrant corners, that is for $\omega>\pi$, solutions do not belong to $H^1(\Omega_f^0)$ since their derivatives tend to infinity. As a consequence, the corresponding ALE mapping may lose its bijectivity. Due to these two problems, the HE technique usually is only able to handle rather small deformations. 

We propose a slight improvement to the HE technique achieved by turning it into an incremental algorithm. Assume that the ALE displacement $\f{u}_a^i$ at time $t_i$ is known. We can use it to deform the initial configuration of the fluid domain $\Omega_f^0$ and obtain the deformed configuration $\Omega_f^i = \Omega_f^0 + \f{u}_a^i$ as equation (\ref{eq:deformDomain}) describes. We can then compute an ALE displacement increment $\delta\f{u}_a^{i+1}$ by solving Laplace's equation in the deformed configuration: 
\begin{align}
\label{eq:ihe1}\Delta\delta\f{u}_a^{i+1} &= \f{0}\text{ in }\Omega_f^i,\\
\delta\f{u}_a^{i+1} &= \f{u}_s^{i+1}-\f{u}_s^i\text{ on }\Gamma^i,\\
\label{eq:ihe2}\delta\f{u}_a^{i+1} &= \f{0} \text{ on }\partial\Omega_f^i\setminus\Gamma^i.
\end{align}
And finally, we define the ALE displacement $\f{u}_a^{i+1}$ at time $t_{i+1}$ as $\f{u}_a^i+\delta\f{u}_a^{i+1}$. Note that the resulting incremental harmonic extension (IHE) technique is not equivalent to the HE technique since ALE increments are computed in deformed configurations of the fluid domain. 

One advantage of the IHE technique is that it uses the ALE displacement from the previous time step. Therefore, IHE can be expected to perform slightly better than the HE technique, meaning that it can handle larger deformations. On the other hand, each step of the IHE technique is formulated in a different configuration of the fluid domain than the previous one. As a result, the technique requires matrix assembly at every time step, which makes it more computationally expensive than HE.

In what follows, we apply the same ideas to other MDTs and consider both their non-incremental and incremental versions, which often share the same advantages and disadvantages as the HE and IHE techniques.

\subsection{Bi-harmonic extension (BE/IBE)}
The HE technique is weak when it comes to large deformations. To overcome this problem, one can search for the ALE displacement as a solution to the bi-harmonic equation:
\begin{align}
\label{eq:biharmeq}\Delta^2\f{u}_a &= \f{0}\text{ in }\Omega_f.
\end{align}
Solutions of the bi-harmonic equation are known to be more regular in comparison to Laplace's equation and do not have problems at reentrant corners \cite{richter2010numerical}.

However, this bi-harmonic extension (BE) technique is often dismissed as too computationally expensive. Indeed, in order to solve the bi-harmonic equation, one has two options: either use $C^1$-conforming elements, which in IGA requires $G^1$-continuity between patches \cite{collin2016analysis,birner2018approximation}; or use mixed elements with an auxiliary variable $\f{q}$ to replace the bi-harmonic equation with two Laplace's equations \cite{boffi2013mixed}: 
\begin{equation}
\Delta\f{u}_a = \Delta\f{q},\sg\Delta\f{q}= \f{0}\text{ in }\Omega_f.
\end{equation}
In this work, we consider only the latter option since it is easier to implement for multi-patch geometries.

In our interpretation, the BE technique has the following form: given the interface displacement $\f{u}_s^i$ at time $t_i$, the ALE displacement $\f{u}_a^i$ is computed by solving the following linear system in the initial configuration of the fluid domain $\Omega_f^0$:
\begin{align}
\label{eq:BE1}\Delta\f{u}_a^i &= \Delta\f{q},\sg \Delta \f{q} = \f{0}\text{ in }\Omega_f^0,\\
\f{u}_a^i &= \f{u}_s^i \text{ on }\Gamma^0,\\
\f{u}_a^i &= \f{0}\text{ on }\partial\Omega_f^0\setminus\Gamma^0,\\
\label{eq:BE2}\nabla\f{q}\cdot\f{n} &= \f{0} \text{ on }\partial\Omega_f^0.
\end{align}

The BE technique shares many similarities with HE which make both techniques very efficient: it does not use information from previous time steps; the linear system has to be assembled only once; the multiple-right-hand-sides approach can be used to compute all displacement components at once. However, the resulting linear system is twice the size of the HE linear system: the matrix is of size $2N\times2N$, and the right-hand size is of size $2N\times d$. Moreover, the linear system has a saddle-point structure, so specialized linear solvers are necessary to solve it efficiently.

Just like with the HE technique, we propose an incremental variation of the bi-harmonic extension (IBE). An increment $\delta\f{u}_a^{i+1}$ is computed by solving the following system in the deformed configuration $\Omega_f^i$:
\begin{align}
\Delta(\delta\f{u}_a^{i+1}) &= \Delta\f{q},\sg \Delta \f{q} = \f{0}\text{ in }\Omega_f^i,\\
\delta\f{u}_a^{i+1} &= \f{u}_s^{i+1}-\f{u}_s^i\text{ on }\Gamma^i,\\
\delta\f{u}_a^{i+1} &= \f{0}\text{ on }\partial\Omega_f^i\setminus\Gamma^i,\\
\nabla\f{q}\cdot\f{n} &= \f{0} \text{ on }\partial\Omega_f^i.
\end{align}
After that, the ALE displacement $\f{u}_a^{i+1}$ at time $t_{i+1}$ is defined as $\f{u}_a^i+\delta\f{u}_a^{i+1}$. 

The IBE technique requires matrix assembly at each time step but can potentially handle larger mesh deformations than the BE technique.

\subsection{Linear elasticity (LE/ILE)}
The next MDT we consider is based on the linear elasticity theory. It is widely used in FSI applications and belongs to the state-of-the-art in the field \cite{bazilevs2013,stein2003mesh,tezduyar2007modelling}. The core idea is to treat the fluid domain as an elastic body and to construct ALE displacement as a solution to the equations of linear elasticity:
\begin{equation}
\dv\pmb{\sigma}_a(\f{u}_a) = \f{0}\text{ in }\Omega_f.
\end{equation}
Here, $\pmb{\sigma}_a$ is the Cauchy stress tensor related to the linearized strain tensor $\pmb{\varepsilon}_a = (\nabla\f{u}_a+\nabla\f{u}_a^T)/2$ by the Hooke's law:
\begin{equation}
\pmb{\sigma}_a = \lambda_a\tr(\pmb{\varepsilon}_a)\f{I} + 2\mu_a\pmb{\varepsilon}_a.
\end{equation}
The Lam\'e parameters $\lambda_a$ and $\mu_a$ depend on Young's modulus $E_a$ and Poisson's ratio $\nu_a$. Since we do not apply volumetric or surface force to the fluid domain, Young's modulus does not affect the resulting ALE displacement. On the other hand, Poisson's ratio is important because it regulates resistance of the fluid mesh to volumetric changes. A too high value (close to 0.5) would result in an almost incompressible behavior, which could lead to excessive distortion of the mesh elements and numerical instabilities. In contrast to that, a too low value (close to 0 or even negative) can reduce resistance of the fluid mesh to bijectivity violation. Therefore, we recommend choosing a value between 0.3 and 0.45.  

Unlike the techniques based on harmonic and bi-harmonic extension, the linear elasticity MDT is best known in its incremental version. That is, given the ALE displacement $\f{u}_a^i$ at time $t_i$, an increment $\delta\f{u}_a^{i+1}$ is computed by solving the linear elasticity equations in the deformed configuration of the fluid domain $\Omega_f^i$:
\begin{align}
\label{eq:ILE1}\dv\pmb{\sigma}_a(\delta\f{u}_a^{i+1}) &= \f{0}\text{ in }\Omega_f^i,\\
\delta\f{u}_a^{i+1} &= \f{u}_s^{i+1}-\f{u}_s^i\text{ on }\Gamma^i,\\
\label{eq:ILE2}\delta\f{u}_a^{i+1} &= \f{0}\text{ on }\partial\Omega_f^i\setminus\Gamma^i.
\end{align}
After that, the ALE displacement $\f{u}_a^{i+1}$ at time $t_{i+1}$ is defined as $\f{u}_a^i+\delta\f{u}_a^{i+1}$. We refer to this technique as incremental linear elasticity (ILE). The ILE technique is known for its robustness and an ability to withstand large mesh deformations. However, little to no research has been conducted to explain its superior behavior.

With respect to computational cost, the ILE technique involves solving a linear system with a matrix of size $dN\times dN$ and a right-hand size of size $dN\times1$. The linear system has to be reassembled at each time step. Note that the size of the linear system depends on a dimension of the problem. Therefore, it scales worse from 2D to 3D than the IHE and IBE techniques.

For the sake of completeness, let us also study a non-incremental version of the ILE technique. The ALE displacement $\f{u}_a^i$ at time $t_i$ is computed by solving the equations of linear elasticity in the initial configuration of the fluid domain $\Omega_f^0$:
\begin{align}
\dv\pmb{\sigma}_a(\f{u}_a^i) &= \f{0}\text{ in }\Omega_f^0,\\
\f{u}_a^i &= \f{u}_s^i\text{ on }\Gamma^0,\\
\f{u}_a^i &= \f{0} \text{ on }\partial\Omega_f^0\setminus\Gamma^0.
\end{align}
We call this the linear elasticity (LE) technique. Although one can only expect it to perform well for small deformations, it is rather computationally inexpensive. Similarly to the HE and BE techniques, the LE technique requires matrix assembly only once and lets one reuse the matrix for every time step.

\subsection{Nonlinear elasticity (TINE)}
The last MDT we present in this paper is based on equations of nonlinear elasticity. The idea is to construct the ALE displacement at each time step as an approximate solution to the local balance equations of linear momentum
\begin{equation}\label{eq:nonlinel_tine}
\dv \f{P}_a(\f{u}_a^i) = \f{0}\text{ in }\Omega_f^0,
\end{equation}
where $\f{P}_a = \f{F}_a\f{S}_a$. To ensure bijectivity of the ALE mapping, we use a logarithmic variation of the neo-Hookean material law
\begin{equation}\label{eq:neohooke}
\f{S}_a = \lambda_a\ln J_a\f{C}_a^{-1} + \mu_a(\f{I}-\f{C}_a^{-1}),
\end{equation}
where $\f{C}_a = \f{F}_a^T\f{F}_a$ is the right Cauchy-Green strain tensor. Similarly to the LE and ILE techniques, the Lam\'e parameters $\lambda_a$ and $\mu_a$ can be computed from Young's modulus $E_a$ and Poisson's ratio $\nu_a$, of which only Poisson's ratio affects the solution of equations (\ref{eq:nonlinel_tine}). 

Due to the term $\ln J_a$  in the neo-Hookean law (\ref{eq:neohooke}), any solution of equations (\ref{eq:nonlinel_tine}) satisfies the bijectivity condition (\ref{eq:bijectivity}). This fact makes the MDT based on equations (\ref{eq:nonlinel_tine}) unique since it explicitly enforces the bijectivity condition. Unfortunately, equations (\ref{eq:nonlinel_tine}) are nonlinear, and an attempt to fully solve them at each time step would make the MDT prohibitively expensive. However, since the ALE mapping should possess certain regularity in time, it is possible to use a solution of equations (\ref{eq:nonlinel_tine}) at time $t_i$ to efficiently construct an approximate solution at time $t_{i+1}$ \cite{shamanskiy2019mesh}. We refer to this technique as tangential incremental nonlinear elasticity (TINE). The TINE technique can be seen as pseudo time-stepping or an example of the continuation methods for nonlinear problems \cite{deuflhard2011newton}.

Let us look under the hood of TINE. It is based on a Newton-like linearization of equations (\ref{eq:nonlinel_tine}). To define it, we need to transform equations (\ref{eq:nonlinel_tine}) into a weak form, also known as variation formulation. To that end, let us define a solution space $\m{V} = (H^1(\Omega_f^0))^d$ and a test space $\m{V}_0 = \{\f{w}\in (H^1(\Omega_f^0))^d\;|\;\f{w}=\f{0}\text{ on }\partial\Omega_f^0\}$. We can then write the weak form of equations (\ref{eq:nonlinel_tine}) as
\begin{align}
\nonumber&\text{find } \f{u}_a\in\m{V}\text{ such that }\forall\f{w}\in\m{V}_0\\
&R(\f{u}_a,\f{w}) = \int\displaylimits_\Omega\f{S}_a:\delta\f{E}_a[\f{w}]d\f{x} = 0.
\end{align} 
Here, $\delta\f{E}_a[\f{w}] = \frac{1}{2}\Big(\f{F}_a^T\nabla\f{w} + \nabla\f{w}^T\f{F}_a\Big)$ is the variation of the Green-Lagrange strain tensor. The Taylor expansion at point $(\f{u}_a,\f{w})$ with an increment $\delta\f{u}_a$ yields
\begin{align}
\nonumber R(\f{u}_a+\delta\f{u}_a,\f{w}) = &R(\f{u}_a,\f{w}) +\\ &DR(\f{u}_a,\f{w})\cdot\delta\f{u}_a + o(||\delta\f{u}_a||), 
\end{align}
where $DR(\f{u}_a,\f{w})\cdot\delta\f{u}_a$ is a directional derivative. We refer to \cite{wriggers2008nonlinear,shamanskiy2019mesh} for details on computing $DR(\f{u}_a,\f{w})\cdot\delta\f{u}_a$. 

The idea of the TINE technique is to use one Newton-like step 
\begin{align}
\nonumber&\text{find } \delta\f{u}_a\in\m{V}\text{ such that }\forall\f{w}\in\m{V}_0\\
&DR(\f{u}_a,\f{w})\cdot\delta\f{u}_a = -R(\f{u}_a,\f{w})
\end{align}
per time step to compute an ALE increment and update the ALE displacement. Concretely, given the ALE displacement $\f{u}_a^i$ at time $t_i$, we find an ALE increment $\delta\f{u}_a^{i+1}$ as a solution of the linear problem 
\begin{align}
DR(\f{u}_a^i,\f{w})\cdot\delta\f{u}_a^{i+1} &= -R(\f{u}_a^i,\f{w}) \sg\forall\f{w}\in\m{V}_0,\\
\delta\f{u}_a^{i+1} &= \f{u}_s^{i+1}-\f{u}_s^i\text{ on }\Gamma^0,\\
\delta\f{u}_a^{i+1} &= \f{0} \text{ on }\partial\Omega_f^0\setminus\Gamma^0. 
\end{align}
After that, we define the ALE displacement $\f{u}_a^{i+1}$ at time $t_{i+1}$ as $\f{u}_a^i+\delta\f{u}_a^{i+1}$.

It is natural to compare the TINE and ILE techniques which are very similar at first glance. Both are incremental techniques based on the elasticity theory; both require solution of a linear system with a matrix of size $dN\times dN$ and a right-hand side of size $dN\times1$; both require matrix assembly at each time step. In general, one can expect both techniques to be roughly equal in computational cost. Unlike ILE, however, the TINE technique explicitly enforces the bijectivity condition (\ref{eq:bijectivity}). Moreover, the TINE technique is based in the initial configuration of the fluid domain. As we show in Sections \ref{chap:testALE} and \ref{chap:testFSI}, this last observation results in crucial differences in behavior of the ILE and TINE techniques when it comes to the accumulated distortion effect.

\subsection{Local stiffening}
Most of the fluid mesh deformation happens along the FSI interface, where the structure displacement is applied as a Dirichlet boundary conditions to the ALE displacement. On the other hand, mesh elements in the vicinity of the stationary boundary of the fluid domain undergo almost no deformation. Therefore, their contribution into processing of the applied interface displacement is negligible. If the deformation could be redistributed away from the FSI interface towards the stationary boundary, the mesh could undergo larger deformations without becoming invalid. This is the idea behind local stiffening, which locally changes the way different elements react to the deformation.

Let $\f{G}:[0,1]^d\to\Omega$ be a parametrization of the computational domain $\Omega$. Imagine that we have to compute integrals corresponding to matrix entries of the discretized linear system. One of the simplest ways to implement local stiffening is to drop the Jacobian determinant $\det\nabla\f{G}$ when transforming the integrals from domain $\Omega$ to parametric domain $[0,1]^d$: 
\begin{equation}\label{eq:simplelocstiff}
\int\displaylimits_\Omega(\cdots)d\f{x} = \int\displaylimits_{[0,1]^d}(\cdots)\det\nabla\f{G}d\pmb{\xi}\to \int\displaylimits_{[0,1]^d}(\cdots)d\pmb{\xi}.
\end{equation}
This local stiffening method was first proposed in \cite{tezduyar1993parallel}. For elasticity problems, (\ref{eq:simplelocstiff}) can be interpreted as a local change of Young's modulus
\begin{equation}\label{eq:fsi:stiffE}
E\to\frac{E}{\det\nabla_{\pmb{\xi}}\f{G}},
\end{equation}
which makes elements with small values of $\det\nabla\f{G}$ stiffer and elements with large values softer. Therefore, the former elements undergo less deformation and are less likely to become invalid.

A more advanced local stiffening method introduced in \cite{stein2003mesh} does not simply drop the Jacobian determinant but changes the degree with which it enters the integrals:
\begin{align}\label{eq:localstiff}
\nonumber\int\displaylimits_\Omega(\cdots)d\f{x} = &\int\displaylimits_{[0,1]^d}(\cdots)\det\nabla\f{G}d\pmb{\xi}\to\\
&\int\displaylimits_{[0,1]^d}(\cdots)(\operatorname{det}\nabla\f{G})^{1-\chi}d\pmb{\xi}.
\end{align}
Here, $\chi\geqslant0$ is called the stiffening degree. The higher the stiffening degree is, the more local stiffening is achieved. $\chi=0$ corresponds to no local stiffening, and $\chi=1$ corresponds to Jacobian dropping (\ref{eq:simplelocstiff}). Too high stiffening degrees, however, may result in excessive mesh distortion. We refer to this method as the Jacobian-based local stiffening.

The Jacobian-based local stiffening acts differently depending on whether the mesh deformation method is formulated in the initial or in the deformed configuration of the fluid domain. Namely, if the integrals for matrix entries are computed in the initial configuration $\Omega_f^0$, the local stiffening is based only on the initial parametrization $\f{G}_f^0$. However, in the case of the deformed configuration $\Omega_f^i$, the local stiffening takes into account already applied deformation since the parametrization $\f{G}_f^i$ of the deformed configuration is defined as $(\f{I}+\f{u}_a^i)\circ\f{G}_f^0$, see equation (\ref{eq:deformDomain}). This effect has both advantages and disadvantages. From one point of view, if a particular mesh element becomes ill-shaped after the deformation, its value of $\det\nabla\f{G}_f^i$ decreases. As a result, this element receives more stiffening, which prevents it from becoming even more ill-shaped or invalid. On the other hand, in case of the ILE technique, this deformation-aware local stiffening essentially makes material properties of the mesh deformation-dependent, which can cause irreversible plastic deformation. For other MDTs based in the deformed configuration, namely IHE and IBE, the effect is similar. As we show in Sections \ref{chap:testALE} and \ref{chap:testFSI}, this irreversible deformation accumulates over time and can significantly affect results of FSI simulations. We refer to this effect as accumulated distortion.

\begin{figure}[H]
	\centering
	\includegraphics[height=4cm]{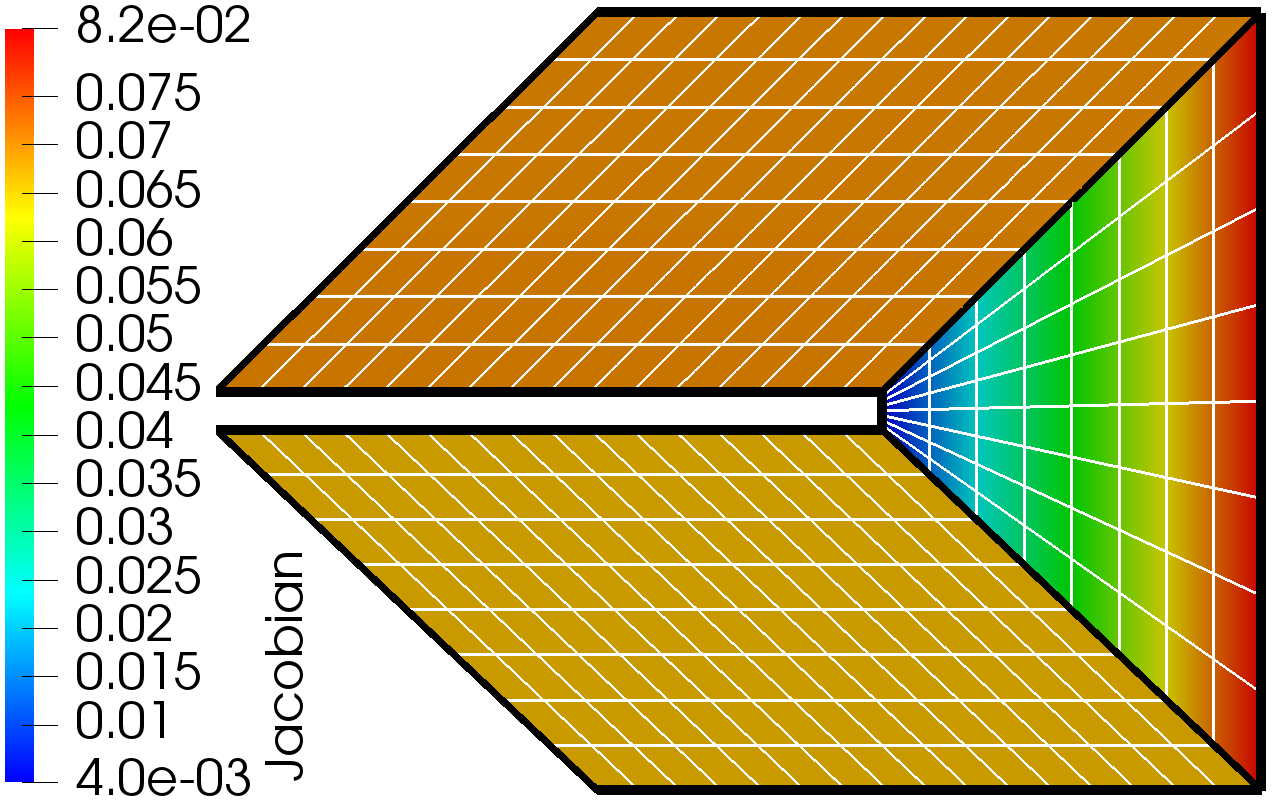}
	\caption{Local stiffening potential of the deforming fluid mesh part.}
	\label{fig:localstiff}
\end{figure}

Regardless if the local stiffening is deformation-aware or not, the initial parametrization $\f{G}_f^0$ of the fluid domain provides a major contribution to how much stiffening each element receives. Let us consider the deforming part of the fluid domain, see Figure \ref{fig:localstiff}. The top and bottom patches are perfect parallelograms, and $\det\nabla\f{G}_f^0$ is constant. Therefore, elements of these patches receive no local stiffening with respect to each other. However, the right patch has a distinct tapered left side, where $\det\nabla\f{G}_f^0$ becomes very small in comparison to the surrounding elements of the top and bottom patches. As a result, element on the left side of the right patch become much stiffer and maintain their shape. Consequently, angles of all three mesh patches that are adjacent to the beam right end do not change much during mesh deformation. In particular, they do not exceed $\pi$, which would lead to $\det\nabla\f{G}_f^i$ becoming negative, which means that the bijectivity condition (\ref{eq:bijectivity}) is not violated.

Note that local stiffening may be harder to achieve in IGA than in classical FEM. We have seen an example of a parallelogram with a natural uniform tensor-product NURBS parametrization. In that case, $\det\nabla\f{G}$ is constant throughout the geometry, and the Jacobian-based local stiffening would have no effect. With a finite element discretization of a parallelogram, it would be possible to place smaller elements where necessary. Since mappings to a reference element are independent for each FEM element, the Jacobian-based local stiffening would have an effect.

In IGA, it is possible to locally control the element size distribution with local refinement, for example by means of THB-splines \cite{vuong2011hierarchical}. However, local refinement does not change the underlying geometry parametrization and values of the Jacobian determinant. Therefore, in order to apply the Jacobian-based local stiffening in IGA, one has two options: either use non-uniform parametrizations with carefully designed knot vectors to artificially construct domain regions with smaller values of the Jacobian determinant; or design patch geometries in a way that naturally defines such domain regions. An example in Figure \ref{fig:localstiff} belongs to the latter approach.

\subsection{Bijectivity check}
Let us briefly discuss ways to check the bijectivity condition (\ref{eq:bijectivity}) in practice. A solution which takes the NURBS nature of the ALE displacement $\f{u}_a$ into account is to express $J_a$ as a NURBS function \cite{Gravesen2014}. If all coefficients in a NURBS representation of $J_a$ are positive, then the displacement $\f{u}_a$ satisfies the bijectivity condition. Unfortunately, this condition is only sufficient and not a necessary one. Therefore, it may often lead to false detection of bijectivity violation. In practice, we resort to a less elegant solution of sampling $J_a$ at the Gaussian quadrature points associated with the NURBS basis of $\f{u}_a$.

Note that whichever method is chosen, it introduces a certain computational overhead to construction of ALE mappings. Nevertheless, we recommend doing some bijectivity check at every time step, or at least with regular intervals. An ALE mapping that does not satisfy the bijectivity condition (\ref{eq:bijectivity}) quickly makes all ensuing computation results meaningless.   

\section{Benchmark ALE: mesh deformation}\label{chap:testALE}
In order to test and compare all mesh deformation techniques (MDTs), we first consider a simplified FSI-like test based on the benchmark introduced in Section \ref{chap:benchmark}. Instead of solving a fully coupled FSI problem, we ignore the fluid component and let the flexible beam oscillate freely in the presence of external acceleration $\f{g}=(0,l)$. We use the beam motion to drive deformation of the three fluid domain patches adjacent to the beam. By varying the parameter $l$, we can regulate magnitude of the mesh deformation. Although this mesh deformation test is artificial, it mimics real deformations occurring in the original benchmark well enough. Moreover, it is significantly less computationally expensive, which allows us to conduct more tests and better assess properties of each MDT.

\begin{figure}[H]
	\centering
	\includegraphics[height=4cm]{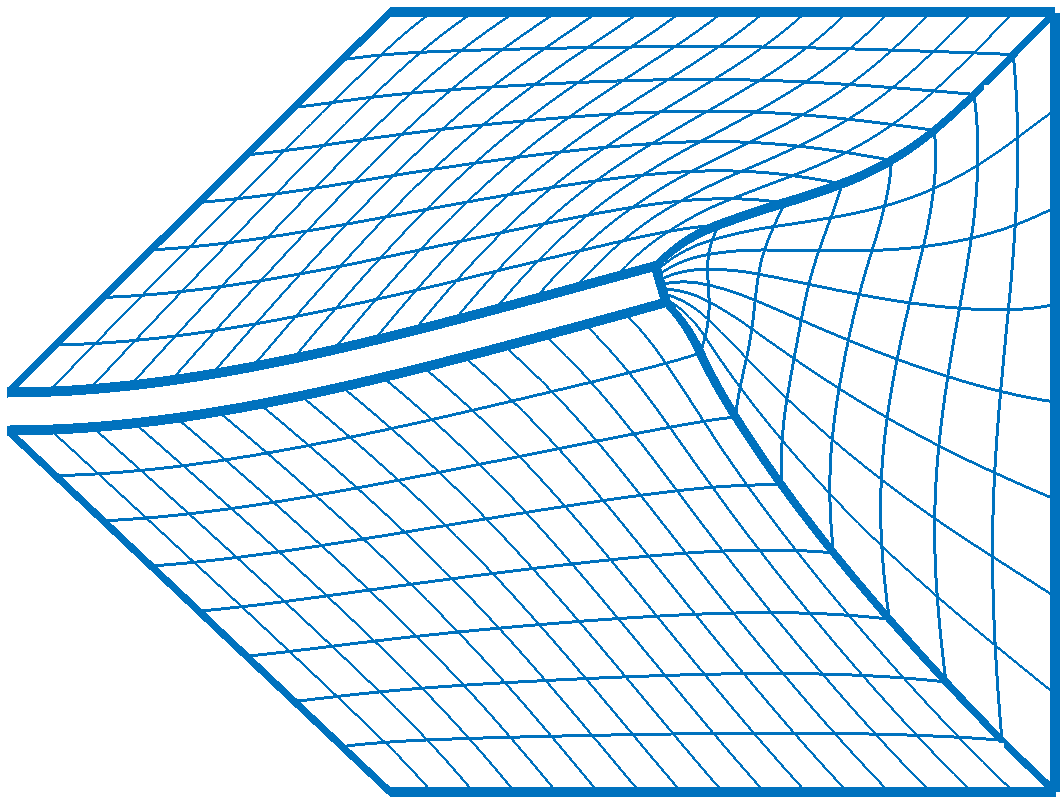}\\
	\vspace{0.1cm}
	\includegraphics[height=4cm]{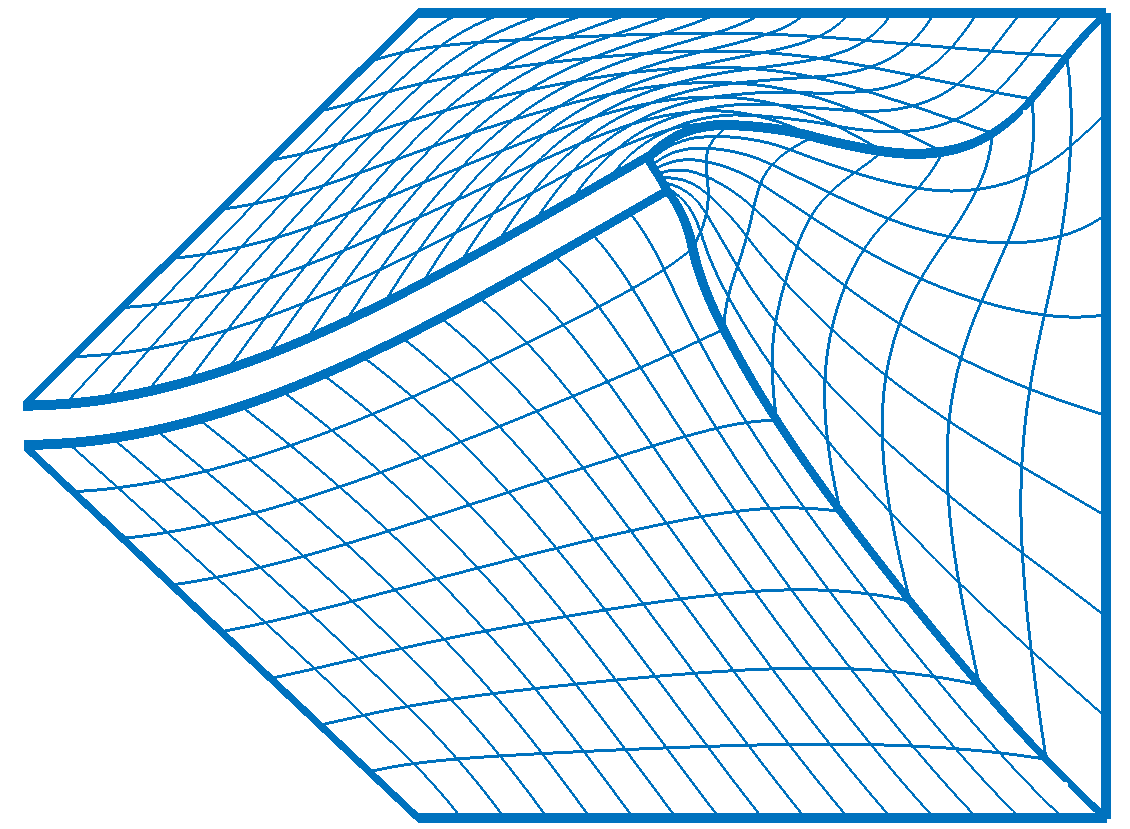}\\
	\vspace{0.1cm}
	\includegraphics[height=4cm]{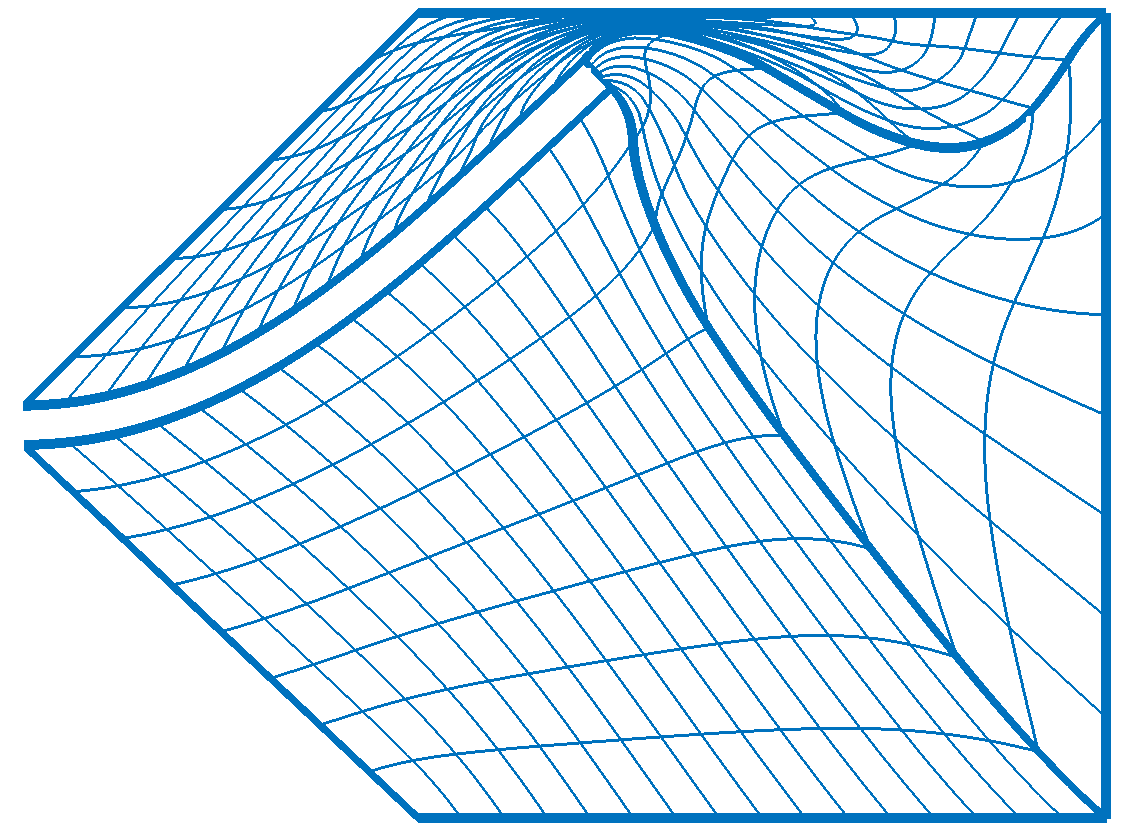}
	\caption{Computational mesh with TINE in the state of maximum deformation for loading levels $l=1,2,3$ (top to bottom).}
	\label{fig:ALElevels}
\end{figure}

For this test, we refine the fluid domain parametrization by applying uniform $h$-refinement thrice. Figure \ref{fig:ALElevels} shows the corresponding computational mesh in the state of maximum deformation for loading levels $l=1,2,3$. We use the following parameters for the structure motion: Young's modulus $E_s=1.4\times10^6$ kg$\cdot$m$^{-1}\cdot$s$^{-2}$, Poisson's ratio $\nu_s=0.4$, density $\rho_s=10^3$ kg$\cdot$m$^{-3}$. For MDTs based on the elasticity theory, we use Poisson's ratio $\nu_a=0.3$. At each time step, we check whether the bijectivity condition (\ref{eq:bijectivity}) holds. We have implemented this test and the original FSI benchmark within G+Smo---an open-source C++ library for isogeometric analysis \cite{jlmmz2014}---using the gsElasticity submodule. As a linear system solver, we use Pardiso---an efficient parallel direct linear solver \cite{pardiso-6.0c}. For reference, all simulations have been performed on a laptop with a 7th generation Intel Core i7 CPU using eight hyper-threads.

\begin{figure}[H]
	\centering
	\includegraphics[height=4cm]{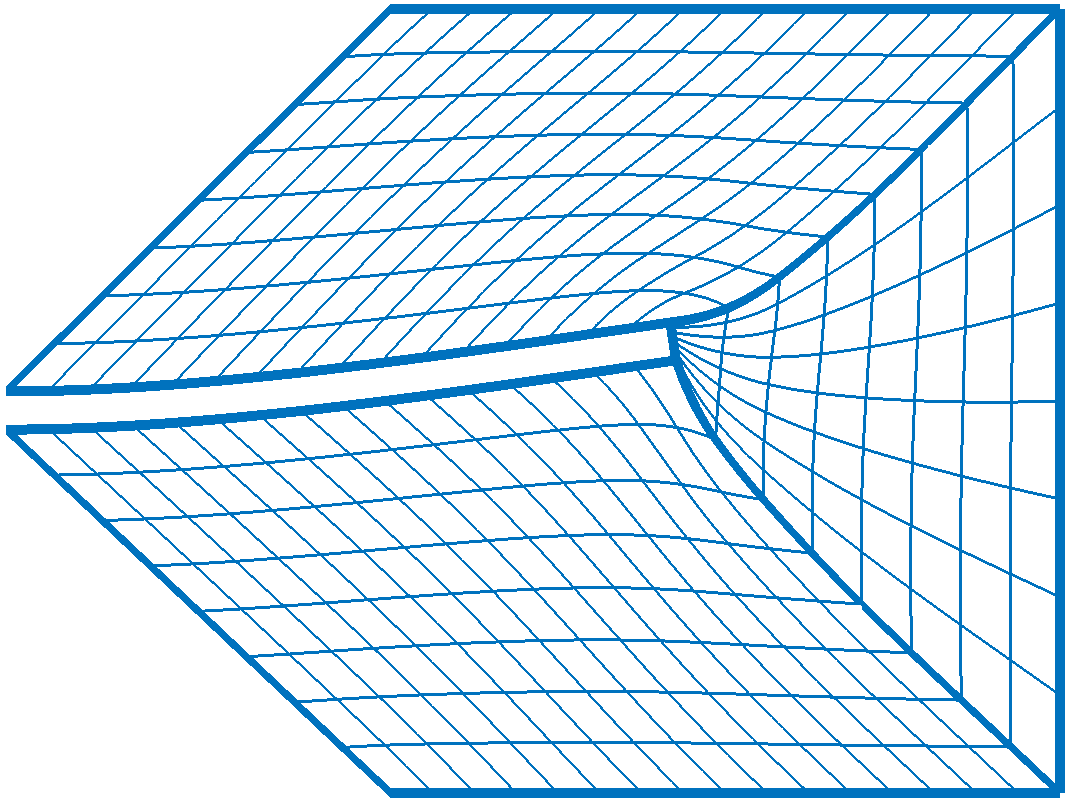}\\
	\vspace{0.1cm}
	\includegraphics[height=4cm]{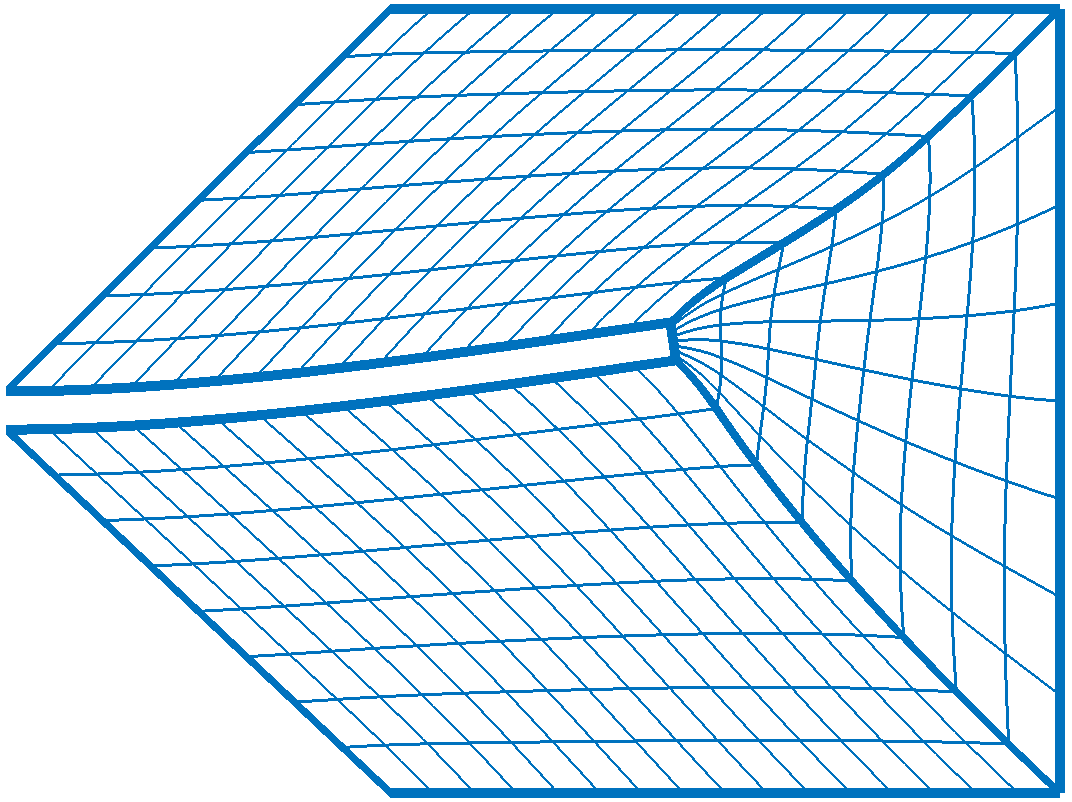}\\
	\vspace{0.1cm}
	\includegraphics[height=4cm]{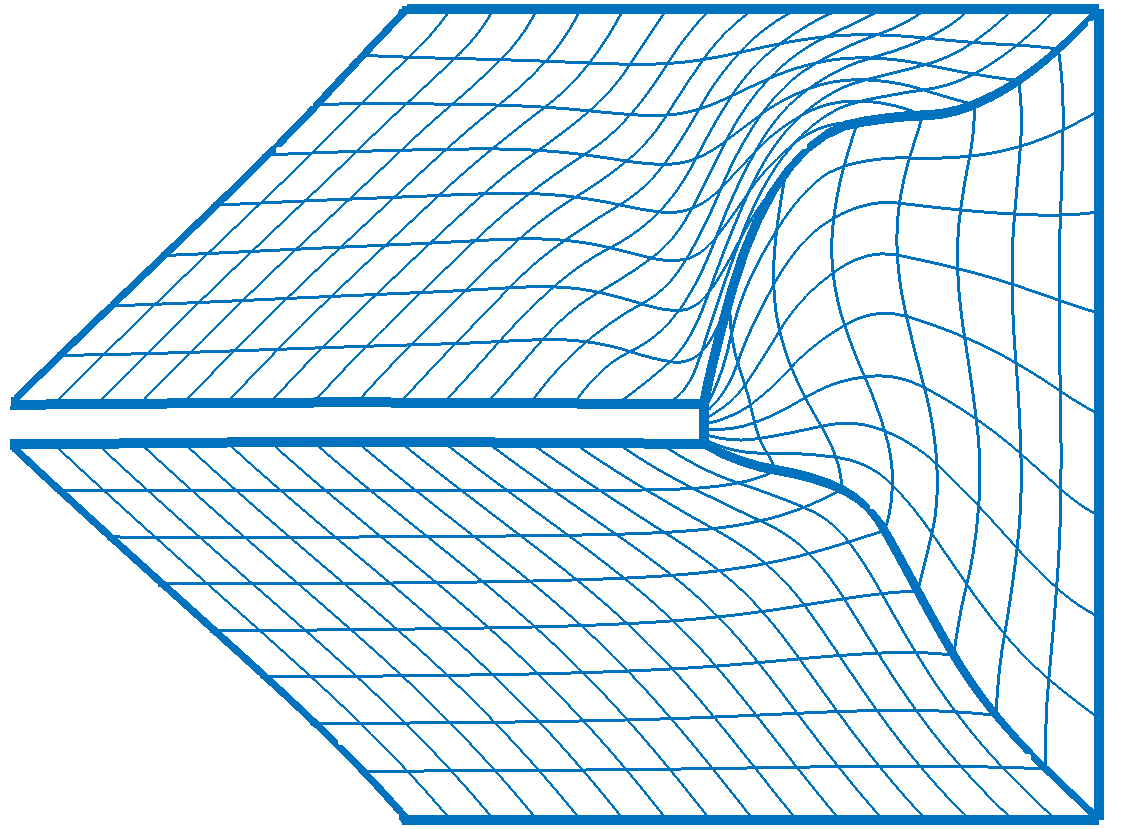}
	\caption{Top: mesh deformation with TINE for loading level $l=0.5$ with stiffening degree $\chi=0$. Middle: mesh deformation with TINE for $l=0.5$ and $\chi=2$. Bottom: accumulated mesh distortion with ILE and $\chi=2$ after 20 oscillation periods with $l=2$.}
	\label{fig:ALEstiff}
\end{figure}

\subsection{Single period test}
First, we consider mesh deformation over one period of beam oscillations. The goal is to study how much deformation each MDT can handle. Here, local stiffening is of crucial importance. Without it, most MDTs can handle only small loading levels $l$. Usually, one of the patch corners adjacent to the right end of the beam becomes larger than $\pi$, which violates the bijectivity conditions (\ref{eq:bijectivity}). With local stiffening, all MDTs can keep these angles below $\pi$ at least for moderate loading levels $l$, see Figure \ref{fig:ALEstiff}. 

\begin{figure}
	\centering
	\includegraphics[height=4cm]{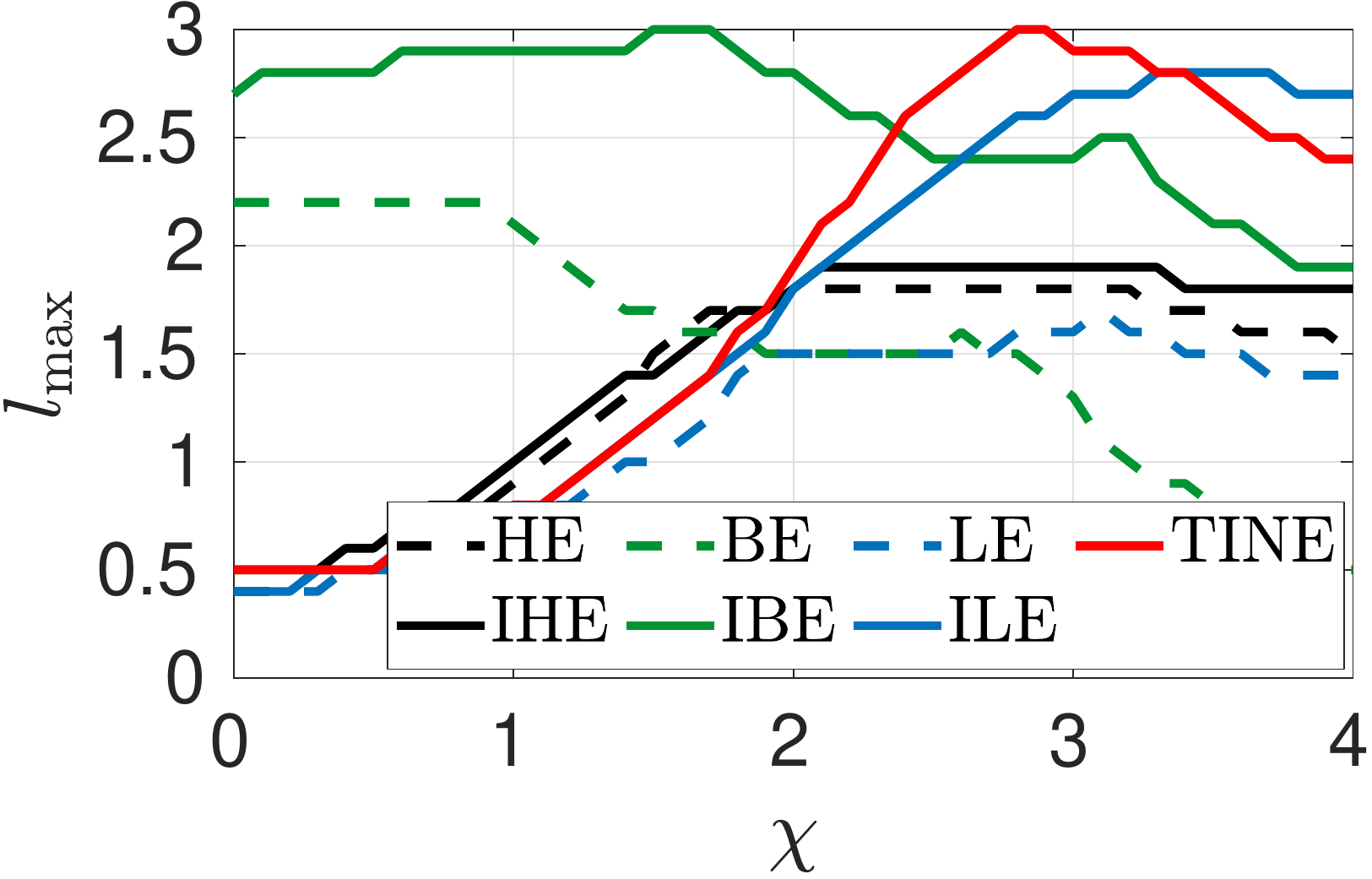}
	\caption{Single oscillation period test: maximal achievable loading level $l_{\max}$ vs the stiffening degree $\chi$ for different MDTs.}
	\label{fig:ALEcomp}
\end{figure}

Figure \ref{fig:ALEcomp} shows a plot of the maximum achievable loading level $l_\text{max}$ versus the stiffening degree $\chi$ for each MDT. Immediately, we can split all MDTs into two groups, with BE and IBE forming one group, and all other techniques belonging to the second group. The main difference between the two groups is that MDTs from the second group can handle almost no deformation without local stiffening. However, as the stiffening degree $\chi$ grows, these MDTs can handle increasingly larger loading levels with maximum loading levels achieved with $\chi\in[2,3]$. With $\chi>3$, we can observe some form of performance deterioration for all MDTs, which is likely caused by too much mesh distortion introduced by the local stiffening.

For the BE and IBE techniques, the behavior is radically different. Already without local stiffening, they can handle larger loading levels than some MDTs from the second group can achieve even with local stiffening. However, as we introduce local stiffening, BE and IBE show almost no response for $\chi < 1$ and start to slowly perform worse for $\chi>1$. 

Overall, MDTs can be ranged with respect to their capability to handle large deformations in the following way: IBE, ILE and TINE are the most powerful and can handle loading levels up of to 2.8--3; BE occupies the second place with the maximum loading level of 2.2; and HE, IHE and LE are the least powerful with maximum loading levels of 1.6--1.9.

\begin{figure}
	\centering
	\includegraphics[height=4cm]{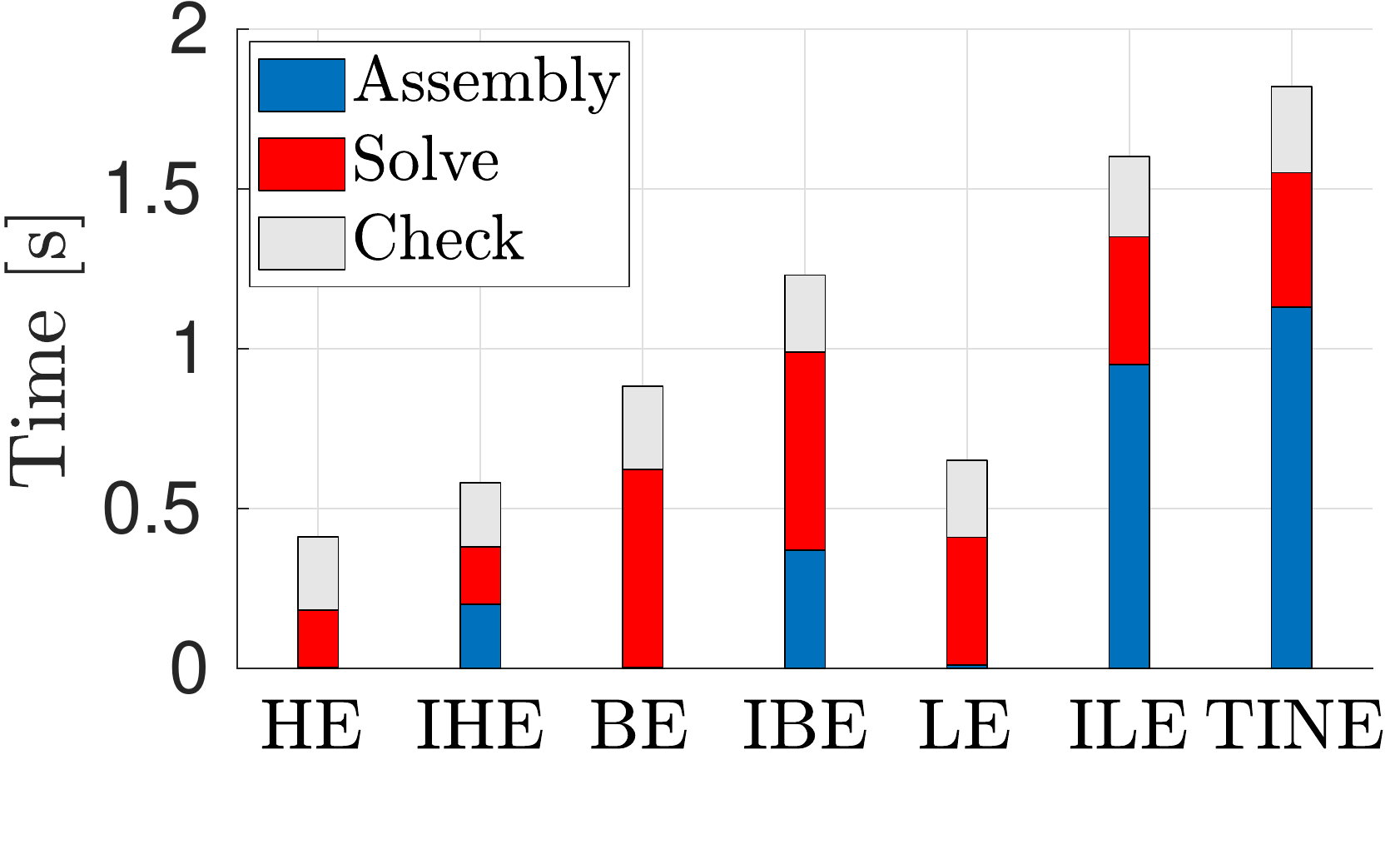}
	\caption{Single oscillation period test: computational time for each MDT split into the assembly, solving and bijectivity check parts.}
	\label{fig:ALEtime}
\end{figure}

Figure \ref{fig:ALEtime} presents an analysis of computational complexity of each MDT. We have measured time required for linear system assembly, linear system solution and an ensuing check of the bijectivity condition (\ref{eq:bijectivity}). Note that this is real time and not CPU time. Although not a perfect measure of algorithm performance, real time still allows us to compare relative efficiency of different MDTs since we have implemented them in the same framework of G+Smo.

The time analysis shows that the HE, BE and LE techniques are significantly faster than their incremental versions. This result is not surprising because non-incremental techniques do not require matrix assembly at each time step. The BE and IBE techniques take the largest amount of time to solve the linear system due to the saddle-point structure of the system. At the same time, relative complexity of the elasticity equations makes the assembly time of the ILE and TINE techniques significantly larger than for other techniques. Overall, the HE, IHE and LE techniques are the most efficient. The ILE and TINE are the most computationally expensive, and the BE and IBE techniques fall in between. All techniques include a small overhead associated with the bijectivity check. This overhead can be reduced by not performing the check at every time step.

\subsection{Long-term behavior test}
\begin{figure}
	\centering
	\includegraphics[height=3.5cm]{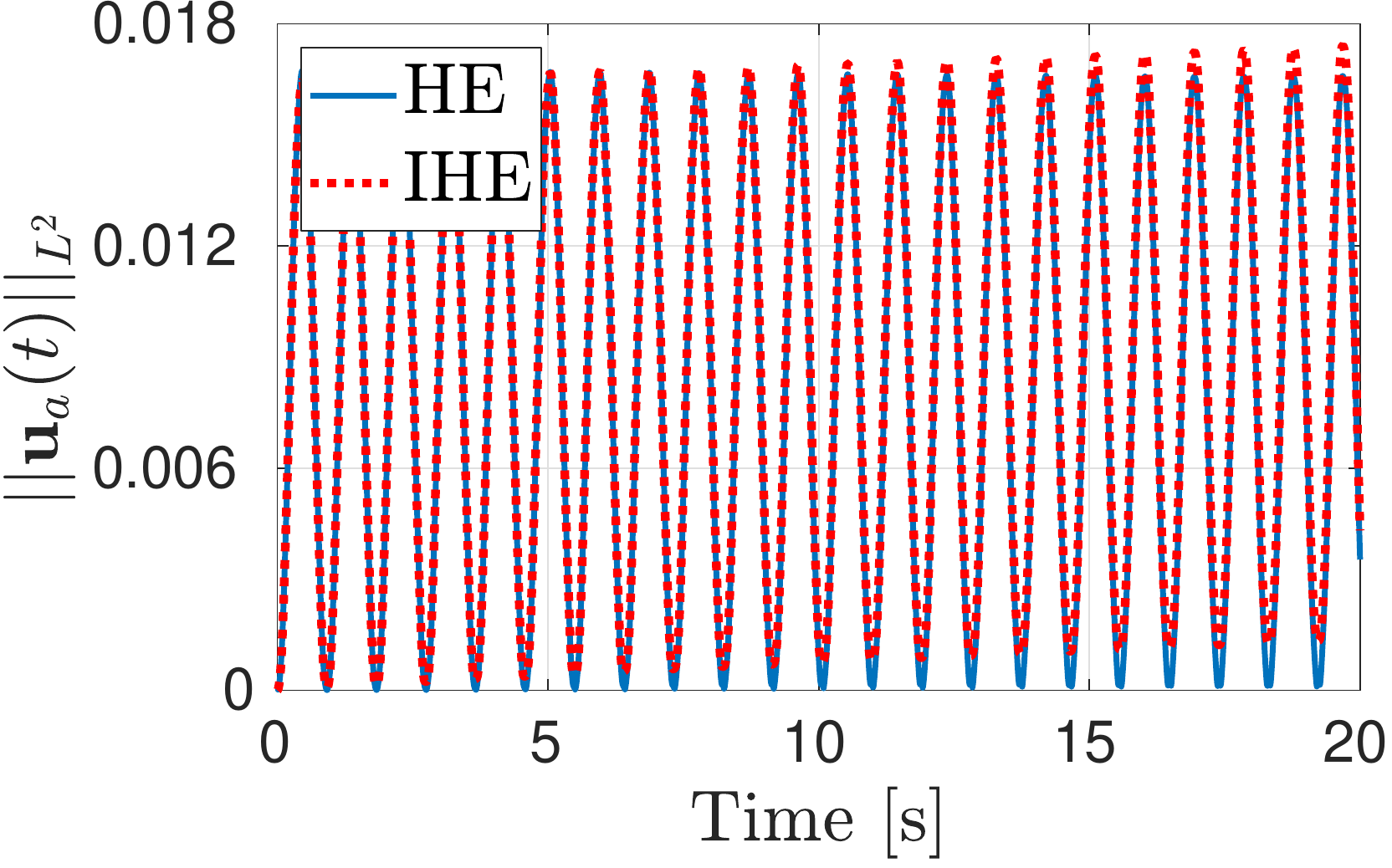}
	\includegraphics[height=3.5cm]{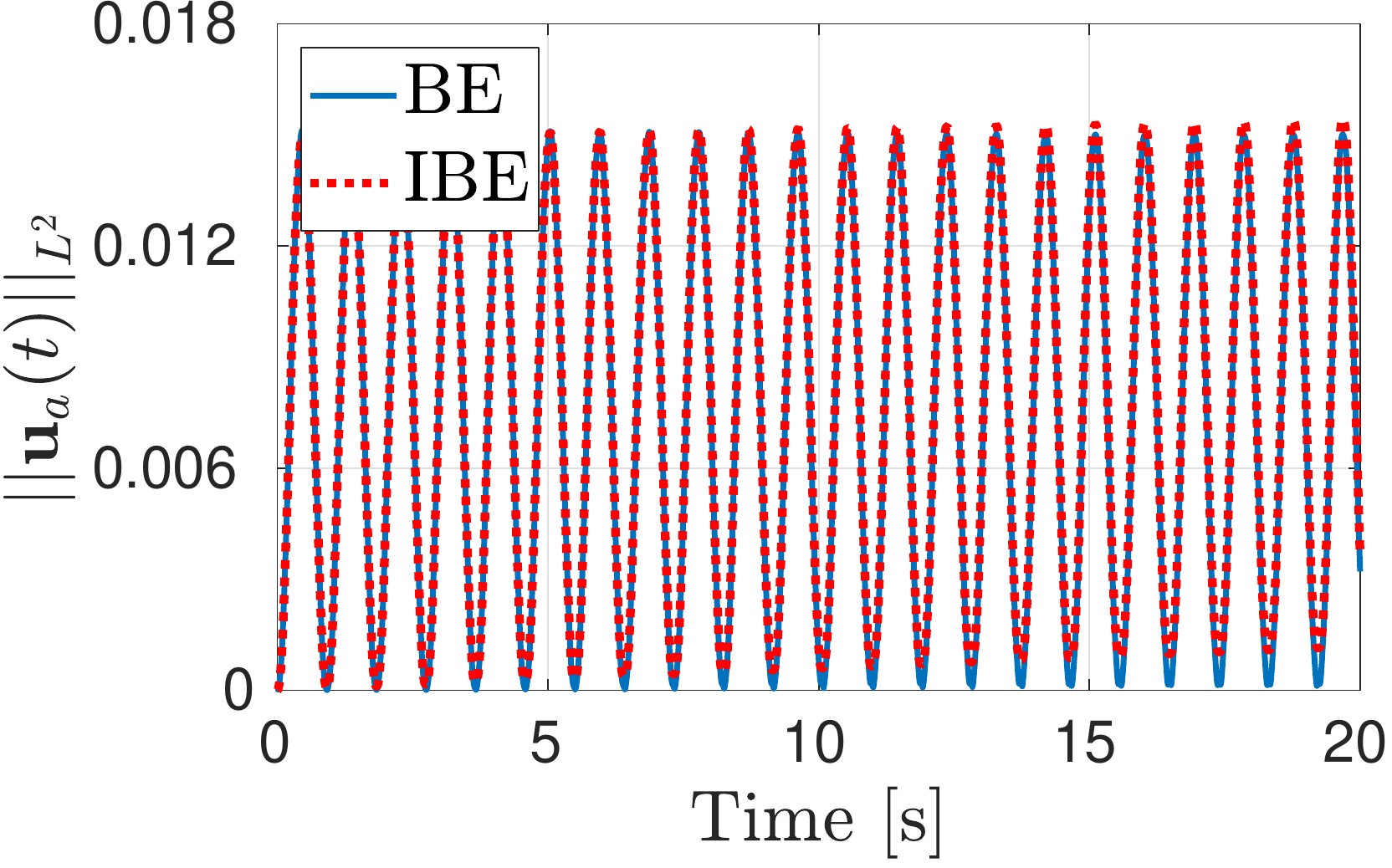}
	\includegraphics[height=3.5cm]{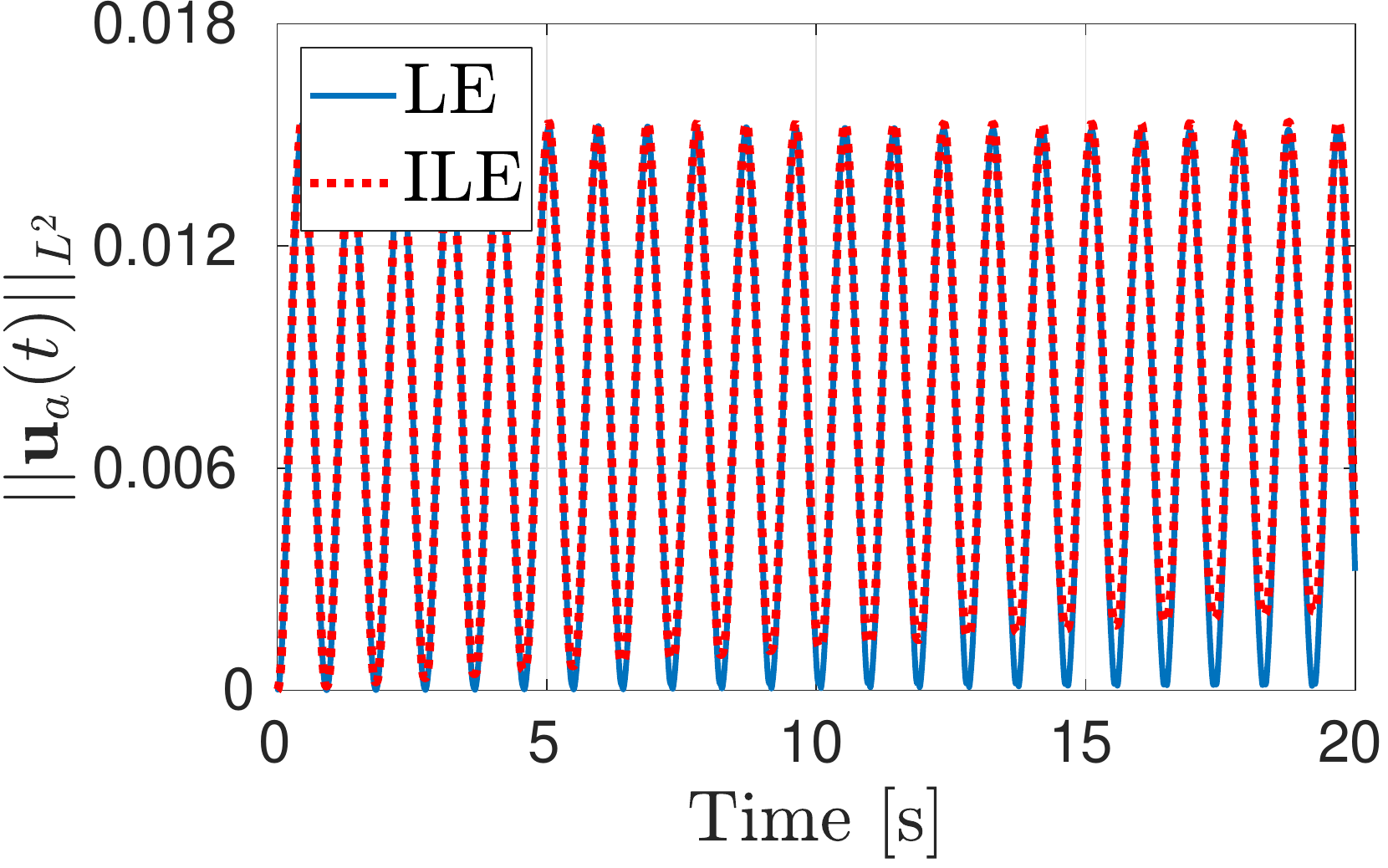}
	\includegraphics[height=3.5cm]{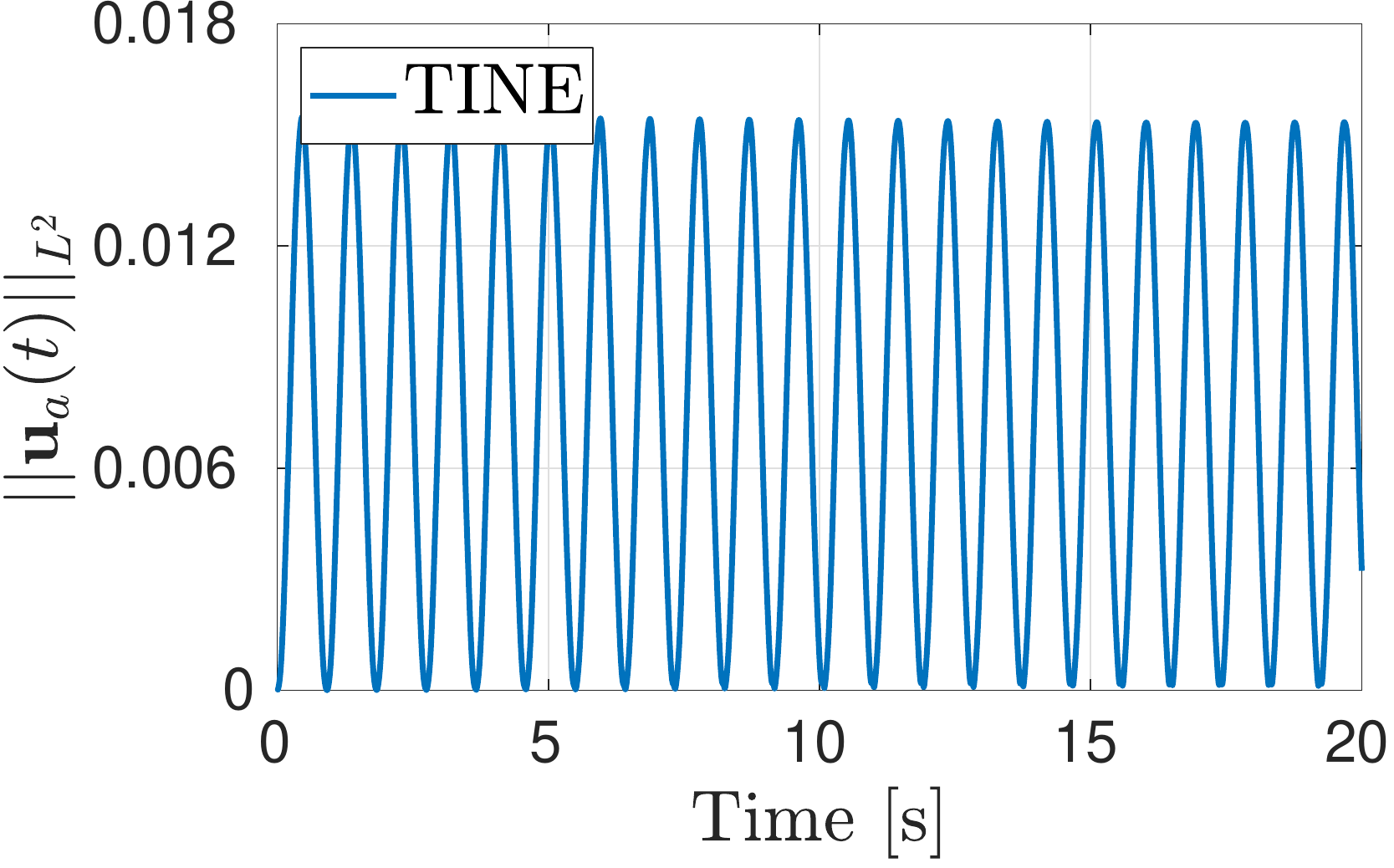}
	\caption{Long-term behavior test. $L^2$-norm of ALE displacement over time.}
	\label{fig:ALEnorm}
\end{figure}

In the second test, we study the long-term behavior of different MDTs and their effect on the fluid mesh. To that end, we perform the simulation over a time period of 20s, which includes roughly 22 periods of beam oscillations. A quantity of interest is the $L^2$-norm of the ALE displacement measured in the initial configuration of the fluid domain. A perfect MDT should return the fluid mesh to its initial state once the beam is not deformed. Therefore, the ALE norm $||\f{u}_a(t)||_{L^2(\Omega_f^0)}$ should be close to zero at the end of each oscillation period. Figure \ref{fig:ALEnorm} shows behavior of the ALE norm over time for each MDT with the loading level $l=1.5$. We have used Jacobian-based local stiffening with $\chi=2$ for all MDTs with the exception of the BE and IBE techniques. These techniques are able to handle the loading level $l=1.5$ without local stiffening.

As Figure \ref{fig:ALEnorm} shows, the ALE norm behaves periodically and returns to zero at the end of each oscillation period with the HE, BE, LE and TINE techniques. On the other hand, with the IHE, IBE and ILE techniques the ALE norm at the end of each oscillation period grows in a monotonous fashion. This effect has been previously reported in \cite{terahara2020ventricle}, and we refer to it as accumulated distortion. It appears only for MDTs which are based in the deformed configuration of the fluid domain. As a result, mesh deformation becomes path-dependent, the fluid mesh does not return to its initial state, and its quality deteriorates over time. Figure \ref{fig:ALEstiff} illustrates the state of the mesh at the end of the simulation with the ILE technique. 

\begin{figure}
	\centering
	\includegraphics[height=3.7cm]{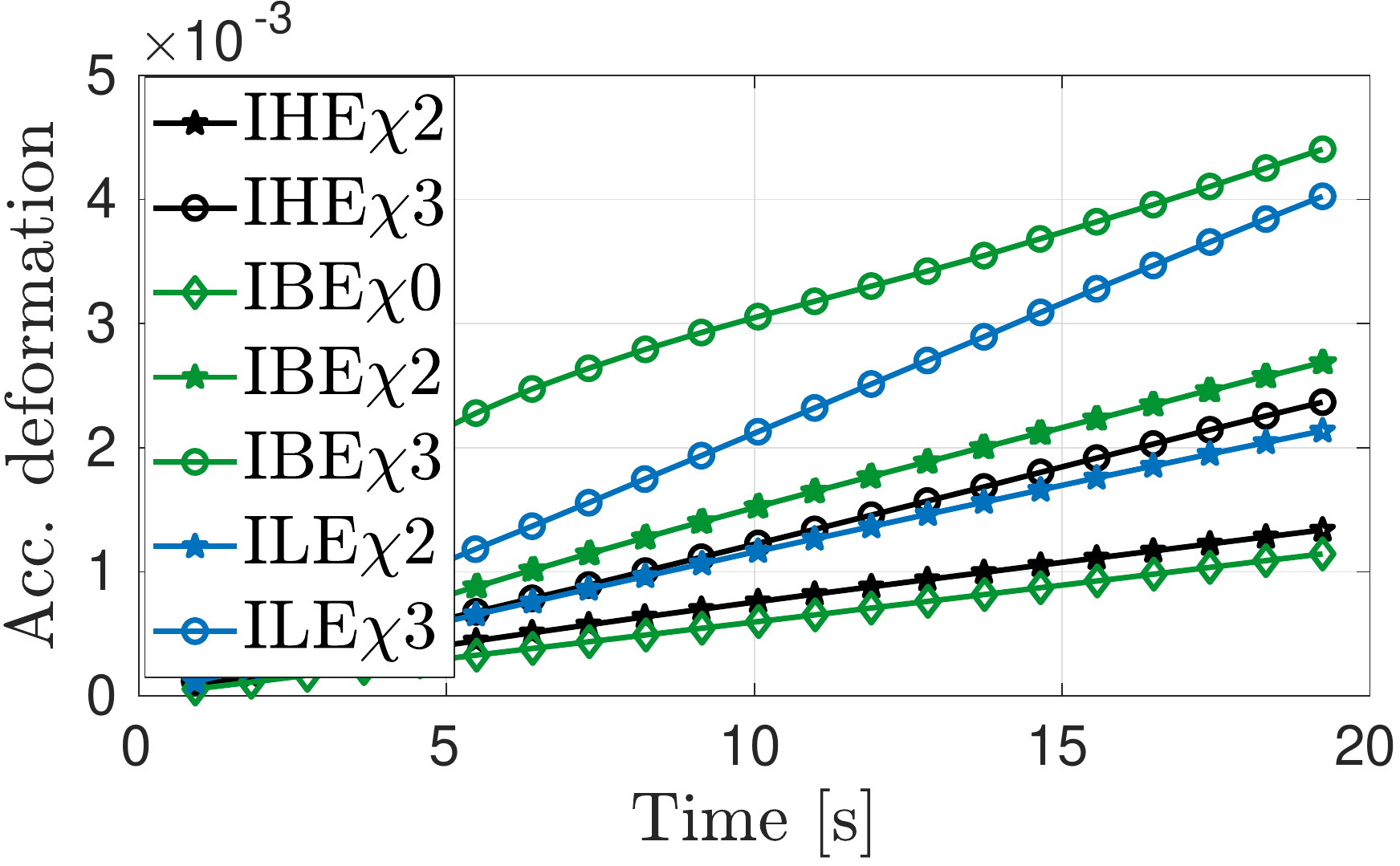}
	\includegraphics[height=3.7cm]{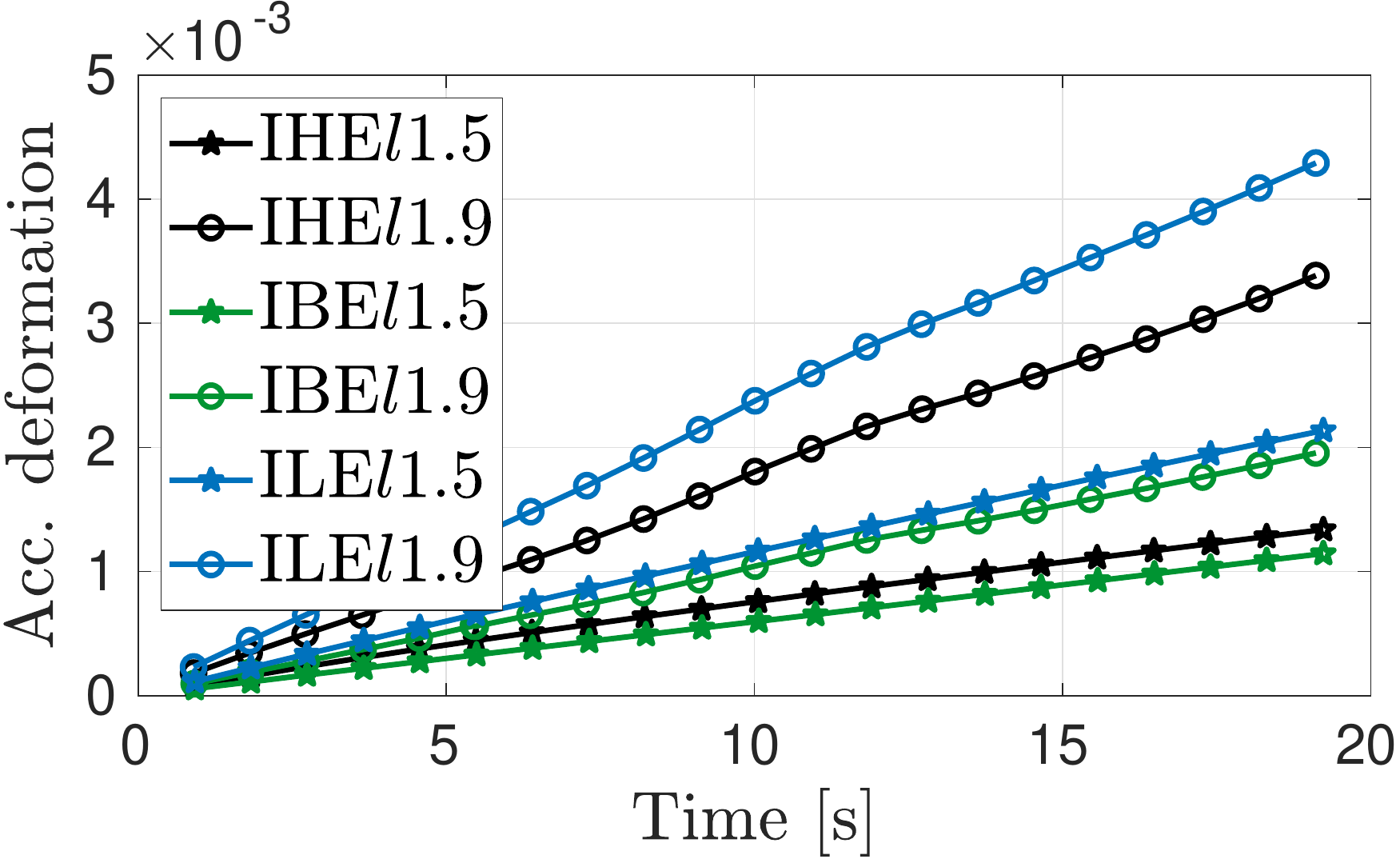}
	\caption{Long-term behavior test. Accumulated distortion for incremental MDTs. Top: fixed loading level $l=1.5$ and varying stiffening degree $\chi$. bottom: fixed stiffening degree $\chi=2$ and varying loading level $l$.}
	\label{fig:accDef}
\end{figure}

The accumulated distortion effect becomes more prominent as the mesh deformation magnitude grows. To study it in more details, we perform the long-term behavior test for the IHE, IBE and ILE techniques with varying values of the loading level and stiffening degree. In Figure \ref{fig:accDef}, we plot values of the ALE norm at the end of each oscillation period. We can observe that both parameters seem to increase the rate of accumulated distortion.

\section{Benchmark FSI2: flow-induced vibrations}\label{chap:testFSI}
In this section, we perform the FSI simulation described in Section \ref{chap:benchmark}. From several simulation scenarios proposed in \cite{turek2006proposal}, we choose a scenario titled FSI2 since it corresponds to the largest magnitude of beam displacement and mesh deformation. Using this FSI simulation, we test and compare different mesh deformation techniques (MDTs) introduced in Section \ref{chap:methods}. We use stiffening degree $\chi=2.5$ for all MDTs but BE and IBE. To them, we apply no local stiffening.

For the analysis, we refine the geometry parametrization five times using uniform $h$-refinement and perform the simulation for 15 seconds with a time step $\Delta t=0.0025$ s. For time integration, we use the Newmark method \cite{wriggers2008nonlinear} with $\beta=0.5$ and $\gamma=1$ for the structure and the IMEX scheme \cite{john2016finite} with $\theta=0.5$ and no stabilization for the fluid. We achieve the coupling of fluid and structure by means of the partitioned Fluid-Dirichlet-Structure-Neumann algorithm \cite{dorfel2011fluid}. 

\begin{figure*}
	\centering
	\includegraphics[height=3.2cm]{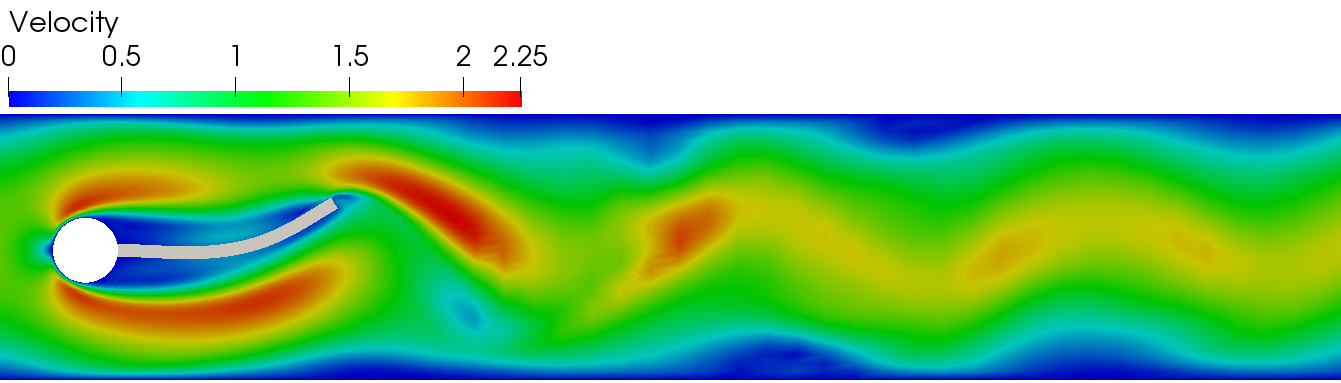}
	\includegraphics[height=3.2cm]{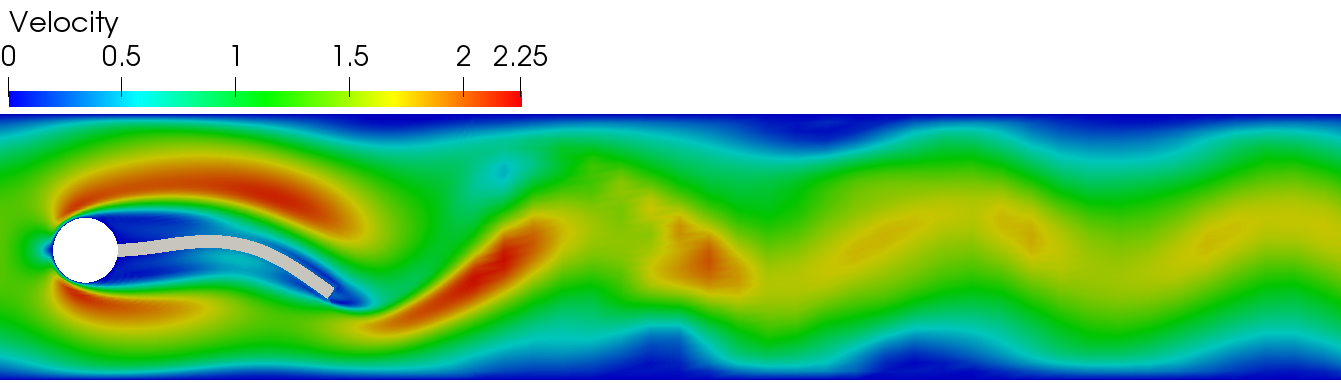}\caption{Benchmark FSI2: fully developed oscillation regime. Fluid velocity field for the maximal upward und downward beam deflection.}
	\label{fig:FSI2velocity}
\end{figure*}

The FSI2 scenario is characterized by the following parameters: fluid density $\rho_f=10^3$ kg$\cdot$m$^{-3}$, fluid kinetic viscosity $\nu_f=10^{-3}$ m$^2\cdot$s$^{-1}$, maximum inflow velocity $v_\text{max}=1.5$ m$\cdot$s$^{-1}$, structure density $\rho_s = 10^4$ kg$\cdot$m$^{-3}$, structure Young's modulus $E=1.4\times10^6$ kg$\cdot$m$^{-1}\cdot$s$^{-2}$, structure Poisson's ratio $\nu_s=0.4$, gravitational acceleration $\f{g}=(0,0)^T$ m$\cdot$s$^{-2}$ and mesh Poisson's ratio $\nu_a=0.3$ (where applicable). The corresponding Reynolds number is $Re = 100$, which results in an unstable flow and development of vortex shedding. Alternating downward and upward forces exerted on the structure by the fluid lead to oscillations of the beam which grow in magnitude until they reach a fully periodic regime. Figure \ref{fig:FSI2velocity} illustrates typical fluid velocity field and beam deformation when the oscillations are fully developed.

\begin{figure*}
	\centering
	\includegraphics[height=3.5cm]{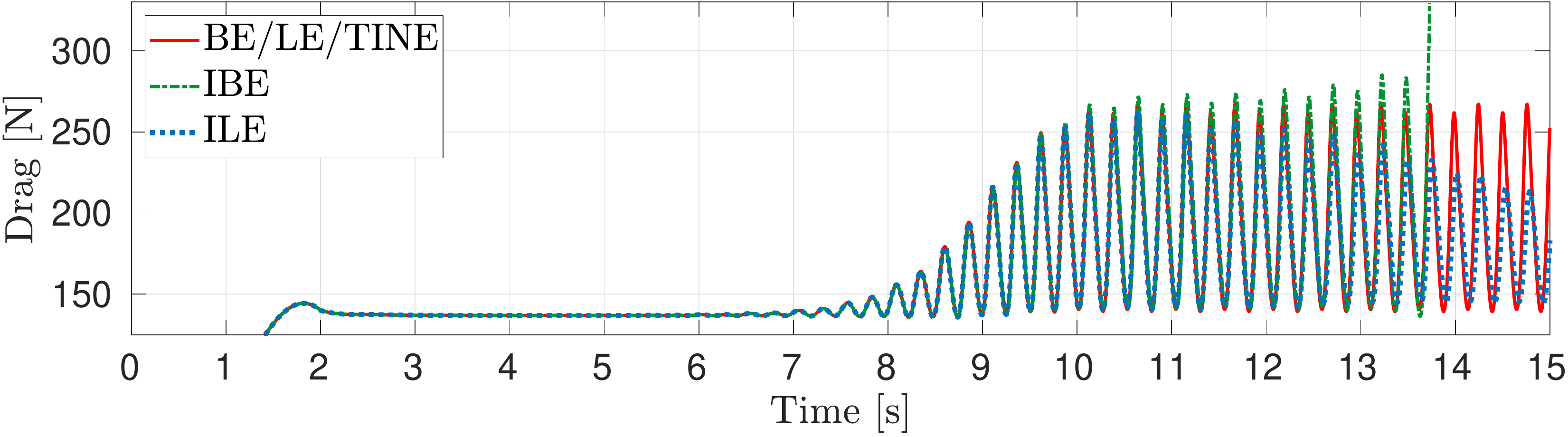}
	\includegraphics[height=3.5cm]{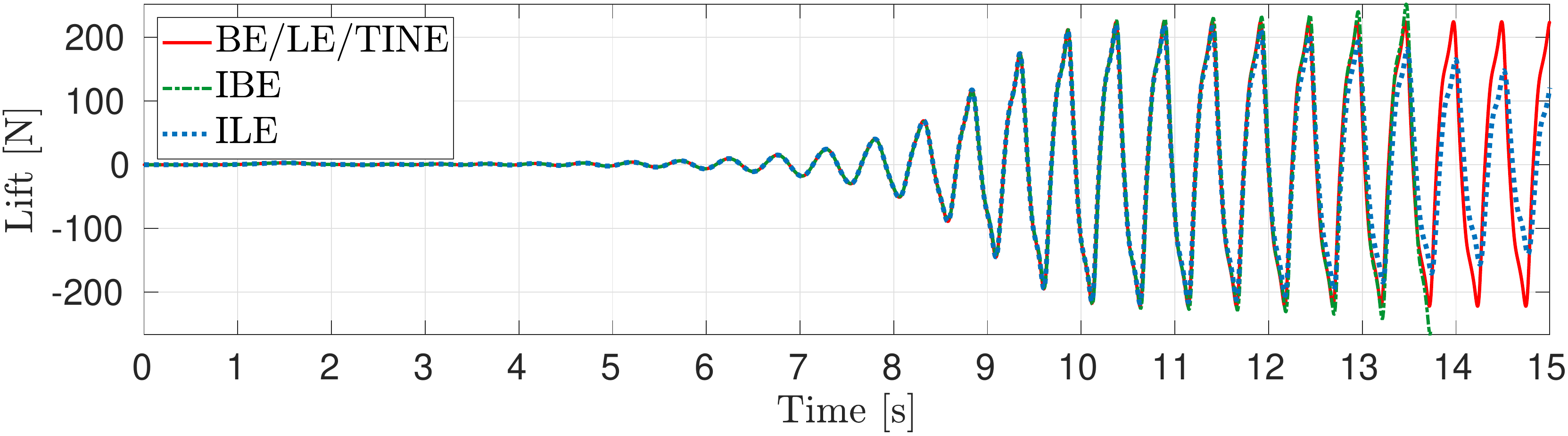}
	\caption{Benchmark FSI2: drag and lift behavior with different MDTs.}
	\label{fig:FSI2_lift}
\end{figure*}

\begin{figure*}
	\centering
	\includegraphics[height=3.5cm]{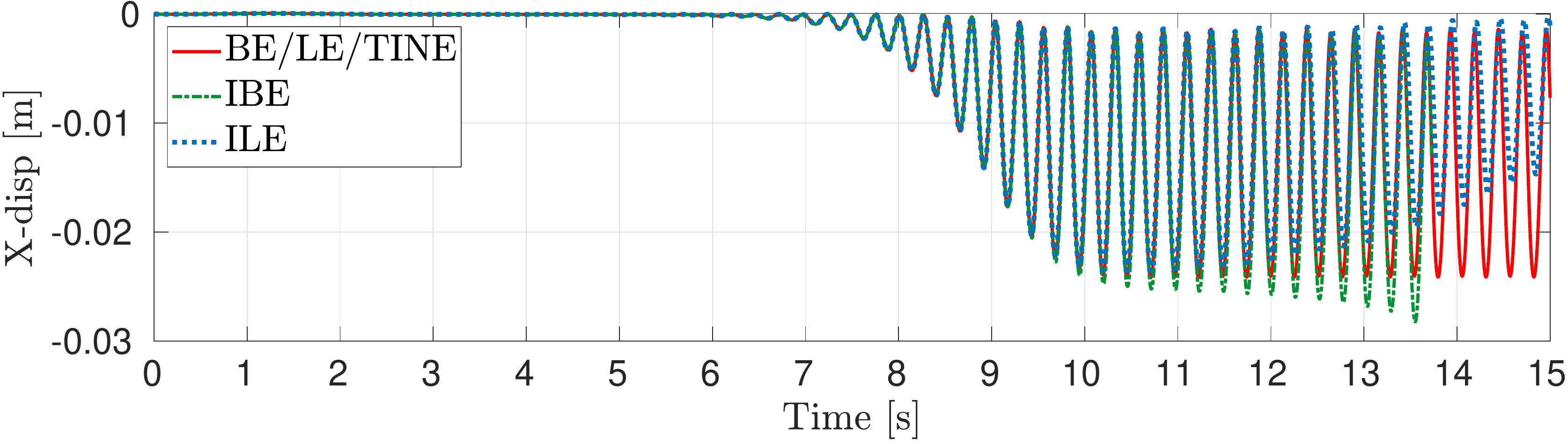}
	\includegraphics[height=3.5cm]{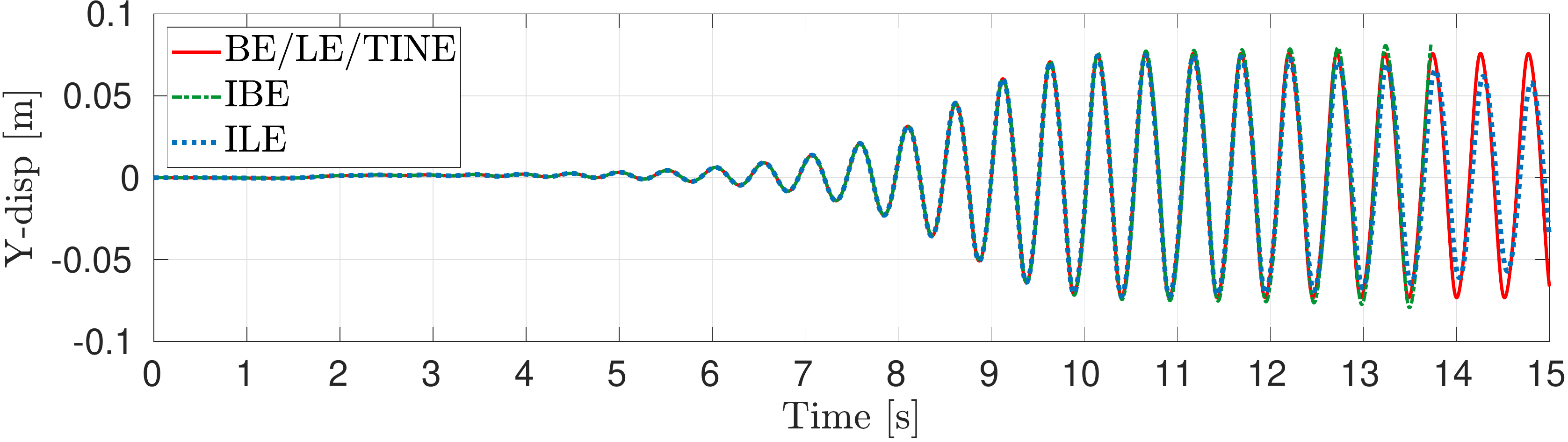}
	\caption{Benchmark FSI2: displacement of the middle point of the beam right end with different MDTs.}
	\label{fig:FSI2_disp}
\end{figure*}

To assess the simulation accuracy, we study the following quantities of interest when the beam oscillations are fully developed : $x$- and $y$-displacement components $u_s^x(A)$ and $u_s^y(A)$ of the point $A$ located in the middle of the beam right end; and drag and lift forces $F_D$ and $F_N$ exerted on the structure by the fluid which are defined as
\begin{equation}
(F_D,F_N) = \int\displaylimits_{\Sigma(t)}\pmb{\sigma}_f(\f{v}_f,p_f)\cdot\f{n}d\f{s}.
\end{equation}
Here, $\Sigma(t)$ is the entire boundary of the submerged solid, including the rigid disk and the flexible beam, at time $t$. Since we expect the quantities of interest to behave periodically, we report them in terms of their mean $\big((*)_\text{max}+(*)_\text{min}\big)/2$, amplitude $\big((*)_\text{max}-(*)_\text{min}\big)/2$ and frequency. In Table \ref{table:FSI2res}, we compare the simulation results obtained with the TINE technique against the reference results from \cite{turek2006proposal}.  Overall, our results seem to undershoot the reference values by about 5$\%$, which can be expected since we use much fewer degrees of freedom and a larger time step than the reference simulation. Despite this discrepancy, we are more than capable of reproducing a qualitatively correct behavior of the system and can use it to study the MDTs. Figures \ref{fig:FSI2_lift} and \ref{fig:FSI2_disp} illustrate behavior of the lift, drag and beam displacement.

\begin{table*}
	\centering
	\caption{Benchmark FSI2: simulation results with TINE ($\Delta t$=0.0025s, 44122 DoFs) vs reference results ($\Delta t$=0.001s, 304128 DoFs).}
	\begin{tabular}{ l c c c c c} 
		\hline\noalign{\smallskip}
		& $F_D$ [$N$]&$F_L$ [$N$] & $u_s^x(A)$ [$10^{-3}m$]&$u_s^y(A)$ [$10^{-3}m$] & Fr [$1/s$] \\
		\noalign{\smallskip}\hline\noalign{\smallskip}
		Results &  203.19$\pm$63.88 & 1.22$\pm$223.35 & -12.75$\pm$11.37 & 1.24$\pm$74.55 & 1.94\\
		Reference &208.83$\pm$73.75& 0.88$\pm$234.2 &-14.58$\pm$12.44& 1.23$\pm$80.6 &2.0 \\ 
		\noalign{\smallskip}\hline
	\end{tabular}
	
	\label{table:FSI2res}
\end{table*}

\begin{figure*}
	\centering
	\includegraphics[height=3.3cm]{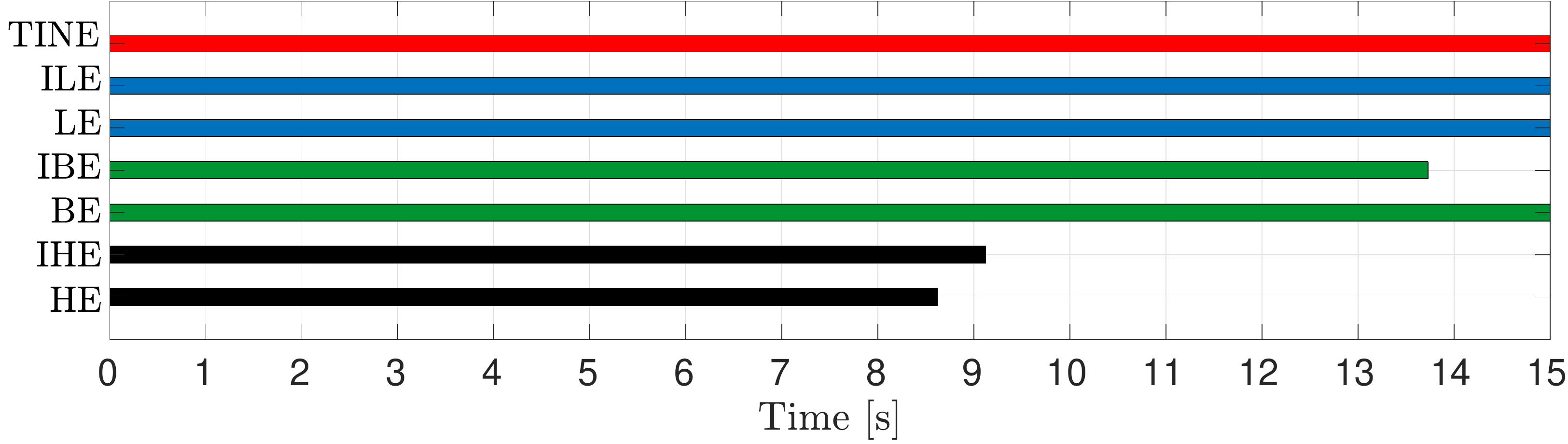}
	\caption{Benchmark FSI2: portion of simulation completed by different MDTs before the ALE mapping becomes invalid.}
	\label{fig:FSI2success}
\end{figure*}

Let us now focus on the fluid mesh deformation. When performing the FSI simulation, we apply each of the seven MDTs (HE, IHE, BE, IBE, LE, ILE and TINE) and study how a particular MDT handles mesh deformations occurring during the simulation. Figure \ref{fig:FSI2success} illustrates what portion of the simulation can be completed using different MDTs. As we can observe, simulations with the HE, IHE and IBE techniques had to be terminated before they could reach the end. All three techniques have failed to maintain bijectivity of the ALE mapping; however, different reasons have led to the failure. In the case of HE and IHE, the simulations have stopped at the ninth second---when the oscillations in the system start to develop. As the applied mesh deformation grows, the HE and IHE techniques fail due to their intrinsic inability to handle large deformations. 

\begin{figure*}
	\centering
	\includegraphics[height=3.3cm]{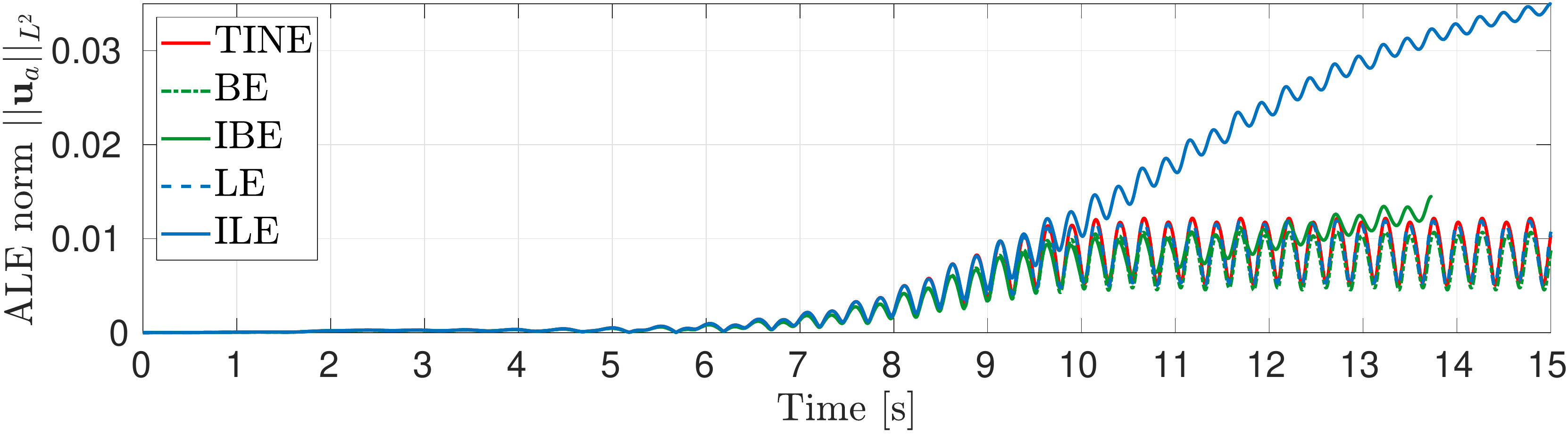}
	\caption{Benchmark FSI2: ALE displacement norm with different MDTs.}
	\label{fig:FSI2_ALEnorm}
\end{figure*}

\begin{figure}
	\centering
	\includegraphics[height=6cm]{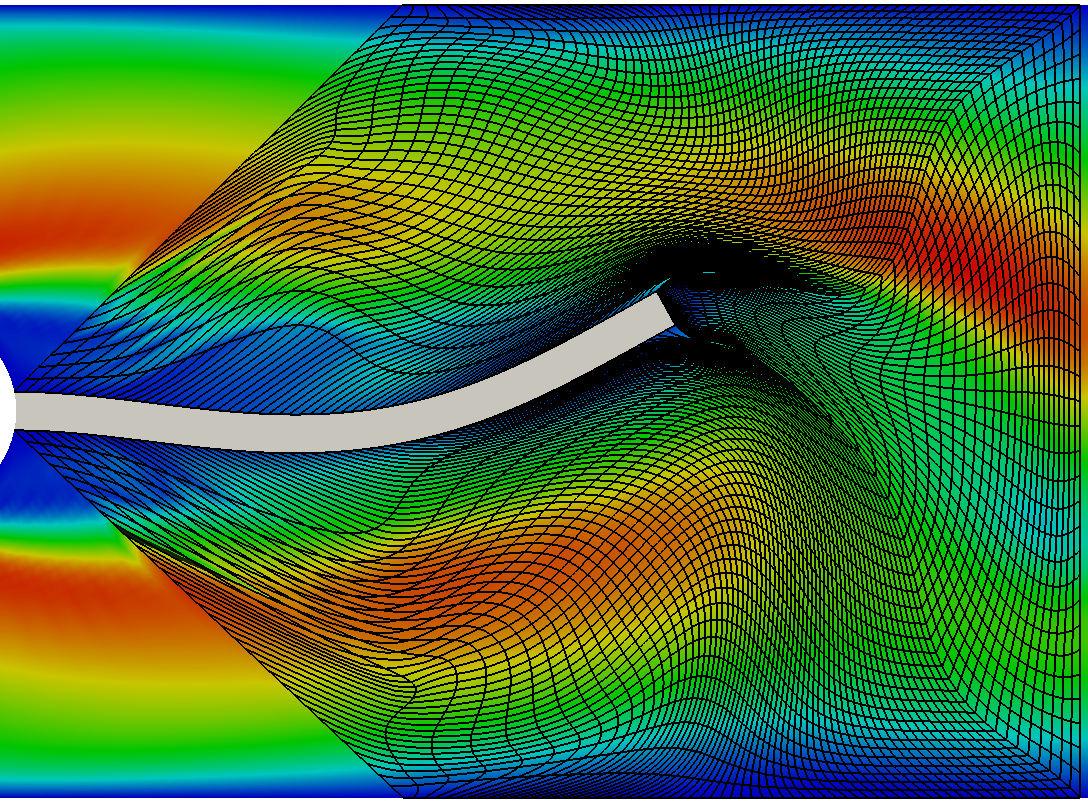}\\
	\vspace{0.2cm}
	\includegraphics[height=6cm]{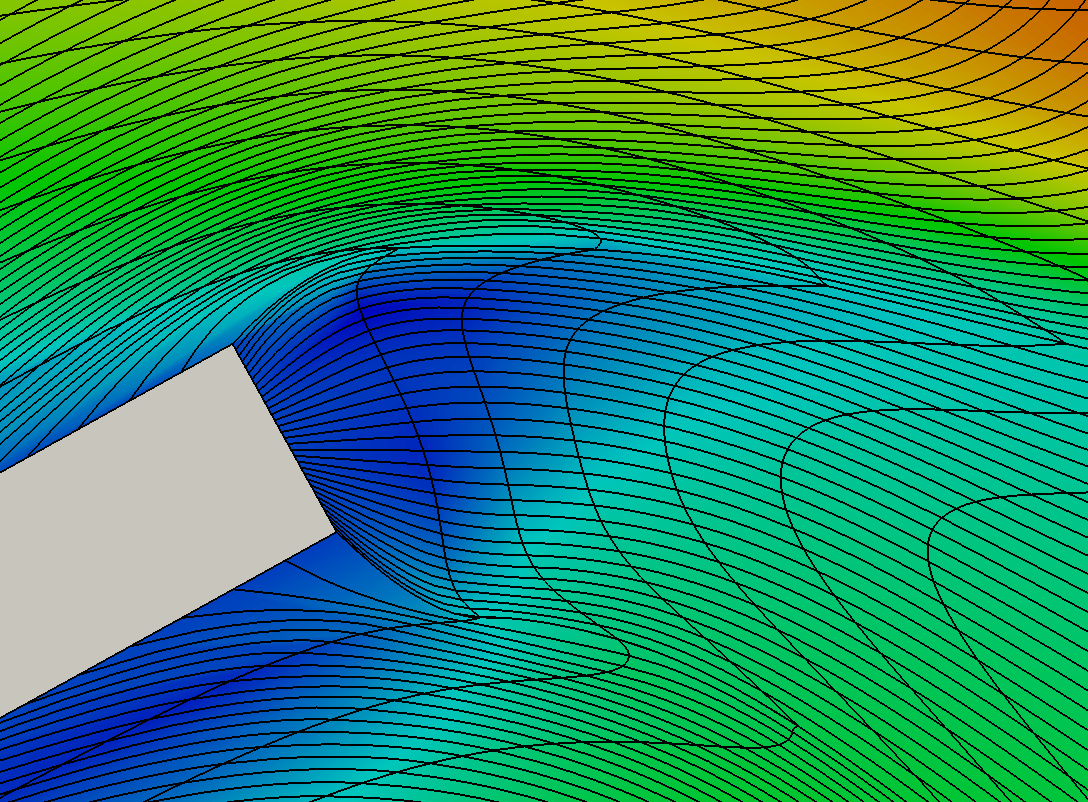}
	\caption{Benchmark FSI2: accumulated distortion of the fluid mesh during the last oscillation period with the ILE technique.}
	\label{fig:FSI2meshILE}
\end{figure}

On the other hand, the IBE technique is able to handle mesh deformations occurring in the simulation but suffers from the accumulated deformation effect described in Section \ref{chap:testALE}. As a result, the simulation fails at the 14th second. Figure \ref{fig:FSI2_ALEnorm} depicts behavior of the ALE norm for all MDTs. We can observe that the ILE technique suffers from even stronger accumulated deformation than IBE but still manages to maintain a bijective ALE mapping until the simulation end. Unfortunately, the highly distorted mesh produced by the ILE technique (see Figure \ref{fig:FSI2meshILE}) affects the simulation results, see Figures \ref{fig:FSI2_lift} and \ref{fig:FSI2_disp}. Instead of a stable periodic behavior in the fully developed oscillation regime, we observe signs of damping. We observe a similar effect when using the IBE technique: the accumulated mesh deformation results in spurious amplification of the oscillations.

\begin{figure}
	\centering
	\includegraphics[height=6cm]{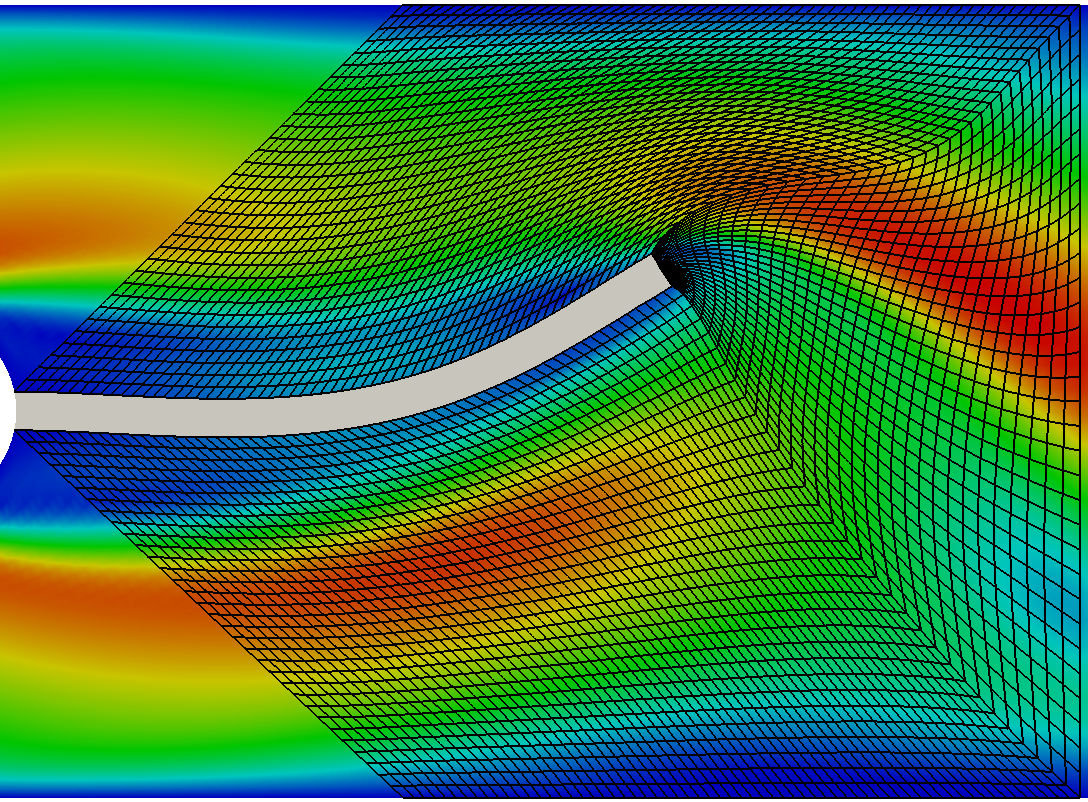}\\
	\vspace{0.2cm}
	\includegraphics[height=6cm]{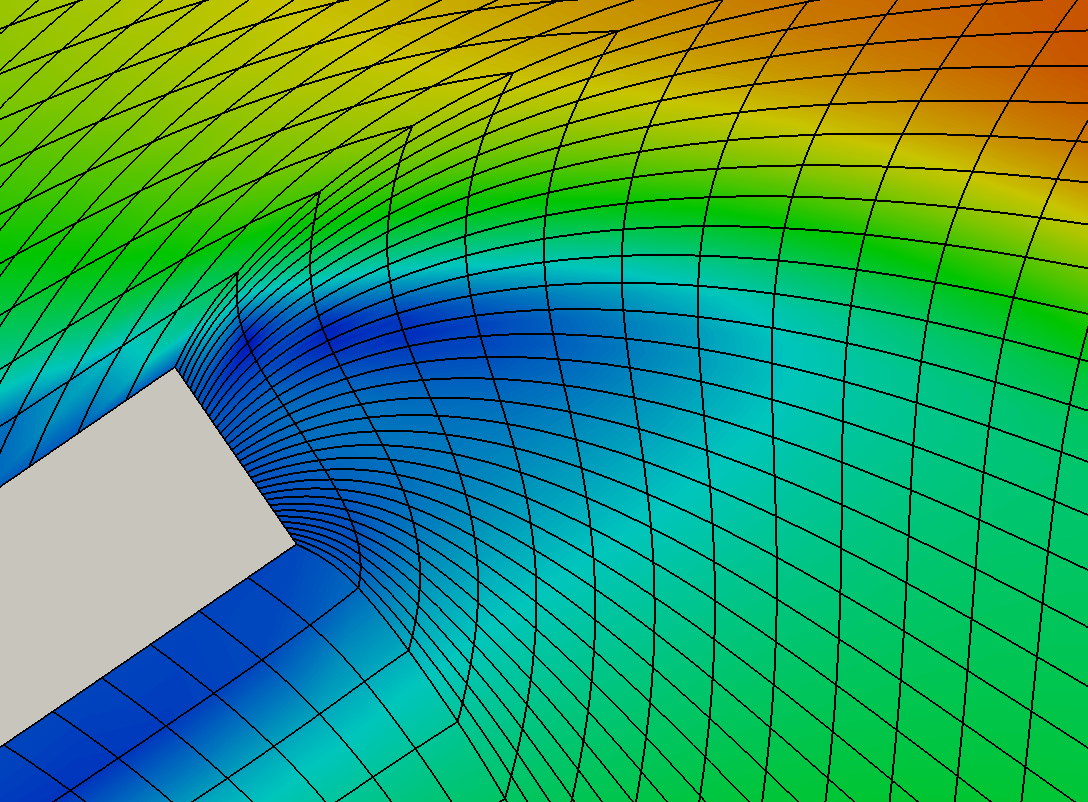}
	\caption{Benchmark FSI2: fluid mesh with during the last oscillation period with the TINE technique.}
	\label{fig:FSI2meshTINE}
\end{figure}

Of the seven MDTs considered in this work, only three---BE, LE and TINE---were able to handle mesh deformations occurring in the simulation and maintain high mesh quality until the simulation end. Most importantly, the BE, LE and TINE techniques have demonstrated no signs of accumulated distortion. Using these techniques, we were able to reproduce a stable periodic behavior of the system and correct simulation results. Figure \ref{fig:FSI2meshTINE} depicts the fluid mesh at the end of the simulation deformed using the TINE technique. Of course, the BE, LE and TINE techniques differ a lot in terms of their computational cost. However, since construction of the ALE mapping corresponds only to a small portion of a total computational effort required to perform an FSI simulation, the choice of MDT does not affect the total computational cost too much. Table \ref{table:FSItimeComparison} compares computational cost of  BE, LE and TINE against the ILE technique which is often considered a default option in the FSI community. 

\begin{table}
	\centering
	\caption{Benchmark FSI2: computational time comparison for MDTs which successfully completed the simulation. The ILE technique time is used as a reference point for comparison.}
	\begin{tabular}{l c c}
		\hline\noalign{\smallskip}
		& ALE time &  Total time\\
		\noalign{\smallskip}\hline\noalign{\smallskip}
		BE   & 2h10m (-48\%) & 23h55m (-7.3\%)\\
		LE & 1h39m (-60\%) & 23h20m (-9.6\%) \\
		ILE  & 4h7m (+0\%) & 25h48m (+0.0\%)\\
		TINE & 4h52m (+18\%) & 26h27m (+2.5\%) \\
		\noalign{\smallskip}\hline
	\end{tabular}
	\label{table:FSItimeComparison}
\end{table}

\section{Discussion and conclusion}\label{chap:concl}
In this work, we have described and compared several mesh deformation techniques (MDTs) which can be used within moving-mesh methods for FSI problems. To evaluate each MDT, we have used a 2D FSI benchmark and its simplified version where the focus lies on mesh deformation. Based on the tests performed in Sections \ref{chap:testALE} and \ref{chap:testFSI}, we can make the following conclusions:
\begin{itemize}
	\setlength\itemsep{-0cm}
	\item Out of seven MDTs that we have considered, two most robust are bi-harmonic extension (BE) and tangential incremental nonlinear elasticity (TINE). Both MDTs can handle large mesh deformations and do not suffer from the accumulated distortion effect.
	\item BE is easier to implement, performs well even without the Jacobian-based local stiffening and is about two times less computationally expensive than TINE. Provided that the saddle-point structure of the linear system is accounted for, we recommend the BE technique as the first method to try in many FSI applications.
	\item The TINE technique is the most computationally expensive of all considered MDTs. However, it can also handle the largest magnitude of mesh deformation when combined with the Jacobian-based local stiffening. Given it high computational cost and implementation complexity, we recommend TINE for FSI applications where extreme mesh deformations are expected.
	\item The incremental bi-harmonic extension and linear elasticity techniques (IBE and ILE) can handle as much mesh deformation as TINE and are slightly less computational expensive. Unfortunately, both techniques suffer from the accumulated distortion effect which can affect the simulation results over long periods of time. We urge the reader to exercise caution when applying this techniques. Detrimental effects of accumulated distortion can be reset by a costly remeshing operation.
	\item Although not suitable for large mesh deformations, the harmonic extension and linear elasticity techniques (HE and LE) can be applied if only small mesh deformations are expected. The exceptional implementation simplicity and low computational cost make HE and LE viable options in certain situations.
	\item Finally, the incremental harmonic extension technique (IHE) can handle only small deformations and suffers from accumulated distortion. We do not recommend using this technique.
\end{itemize}
We would like to emphasize that the performance of all MDTs is dependent on the chosen parametrization of the fluid domain. However, it is unlikely that the MDT behavior will be qualitatively different if a different geometry parametrization is used. 

With respect to the further research directions, we see the following possibilities. One could study the effect of different geometry parametrizations on the MDT behavior. Moreover, isogeometric methods in FSI could benefit from alternative local stiffening approaches. The commonly used Jacobian-based local stiffening has no effect if a uniform geometry parametrization is used. Finally, it would be interesting to combine BE and TINE with additional augmentation techniques such as the solid layer extension and mesh element relaxation.

\begin{acknowledgements}
We would like to express our gratitude to Dr. Michael Helmut Gfrerer and Dr. Matthias M\"{o}ller for extensive and fruitful discussions about fluid-structure interaction. The support of this research by the German Research Council (DFG) under Grant No. SI 765/5-1 (project YASON) and by the German Federal Ministry of Education and Research (BMBF) under Grant No. 05M16UKD (project DYMARA) is greatly acknowledged. 
\end{acknowledgements}

\section*{Conflict of interest}
The authors declare that they have no conflict of interest.

\bibliographystyle{spmpsci}    
\bibliography{refs}

\end{document}